\documentstyle[11pt,epsfig]{article}
\def\la{\langle}\def\ra{\rangle}
\def\be{\begin{eqnarray}}\def\bea{\begin{eqnarray}}
\def\ba{\begin{eqnarray}}
\def\ee{\end{eqnarray}}\def\eea{\end{eqnarray}}
\def\ea{\end{eqnarray}}
\def\ben{\begin{eqnarray}}\def\bitem{\begin{itemize}}
\def\een{\end{eqnarray}}\def\eitem{\end{itemize}}
\def\del{\partial}
\def\G0p{$G_0^\prime$}
\def\ie{{\it i.e}}\def\bi{\bibitem}

\def\N1520{$N^\star (1520)$}
\def\calL{{\cal L}}\def\calF{{\cal F}}\def\calA{{\cal A}}
\def\calO{{\cal O}}\def\calF{{\cal F}}
\def\calD{{\cal D}}\def\calV{{\cal V}}\def\M{{\cal M}}
\def\calM{{\cal M}}
\def\prl{Phys. Rev. Lett.}\def\pr{Phys. Rev.}\def\np{Nucl. Phys.}
\def\pl{Phys. Lett.}
\def\B#1{{}^{#1}\mbox{B}}

\def\Tr{{\mbox{Tr}}}
\newcommand{\e}{{\mbox{e}}}\def\del{\partial}

\def\roughly#1{\mathrel{\raise.3ex\hbox{$#1$\kern-.75em%
\lower1ex\hbox{$\sim$}}}}\def\lsim{\roughly<}
\def\gsim{\roughly>}

\def\calV{{\cal V}}
\def\B#1{{}^{#1}\mbox{B}}

\def\itt{\indent\indent}

\def\B0{\mbox{\boldmath $0$}}
\def\bLambda{\bar{\Lambda}}
\def\bq{\begin{equation}}
\def\eq{\end{equation}}
\def\etap{\eta^\prime}
\def\b3{\mbox{\boldmath $3$}}\def\b6{\mbox{\boldmath $6$}}
\def\hO{\hat{\Omega}}

\renewcommand{\thefootnote}{\fnsymbol{footnote}}
\begin{document}
\begin{titlepage}
\begin{center}

 \vskip 1.0cm
{\Large \bf Lectures on Effective Field Theories for} \vskip 0.1cm

{\Large\bf Nuclei, Nuclear Matter and Dense
Matter}~\footnote{Based on lectures given at the 10th Taiwan
Nuclear Physics Spring School, Hualien, Taiwan, 23-26 January
2002.}
  \vskip 2.2cm
   {{\large Mannque Rho} }
 \vskip 0.2cm

 {\it  Service de Physique Th\'eorique, CEA Saclay, 91191
Gif-sur-Yvette, France}

{and}

{\it  School of Physics, Korea Institute for Advanced Study, Seoul
130-012, Korea}

\end{center}

\vskip 0.5cm

\centerline{(\today)}
 \vskip 3cm

\centerline{\bf Abstract}
 \vskip 0.5cm
This note is based on four lectures that I gave at the 10th Taiwan
Nuclear Spring School held at Hualien, Taiwan in January 2002. It
aims to correlate the old notion of Cheshire Cat Principle
developed for elementary baryons to the modern notion of
quark-baryon and gluon-meson ``continuities" or ``dualities" in
dilute and dense many-body systems and predict what would happen
to mesons when squeezed by nuclear matter to high density as
possibly realized in compact stars. Using color-flavor locking in
QCD, the vector mesons observed at low density can be described as
the Higgsed gluons dressed by cloud of collective modes, i.e.,
pions just as they are in superdense matter, thus showing the
equivalence between hidden $flavor$ gauge symmetry and explicit
$color$ gauge symmetry. Instead of going into details of
well-established facts, I focus on a variety of novel ideas --
some solid and some less -- that could be confirmed or ruled out
in the near future.
\end{titlepage}
\newpage
\tableofcontents
\newpage
\renewcommand{\thefootnote}{\arabic{footnote}}
\setcounter{footnote}{0}

\section{INTRODUCTION}\label{intro}
\setcounter{equation}{0} 
\renewcommand{\theequation}{\mbox{\ref{intro}.\arabic{equation}}}
\itt Let me begin with the question why do we need effective field
theory (EFT in short) for nuclear physics? After all, there is
quantum chromodynamics (QCD) with quarks and gluons figuring as
microscopic degrees of freedom that is supposed to answer all the
questions in the strong interactions, so why not just do QCD, the
fundamental theory? The answer that everybody knows by now is  a
clich\'e: Nuclear physics involves the nonperturbative regime of
QCD and cannot be accessed by a controlled method known up to date
given in terms of the microscopic variables.

Next granted that EFT does mimic QCD in the nonperturbative
regime, what then is the objective of doing effective field
theories in nuclear physics?

In my opinion, there are, broadly speaking. two reasons for doing
it. One is the obvious one often invoked: It is to confirm that
nuclear physics is $indeed$ an integral part of the Standard
Model, namely its strong interaction component (i.e., the
$SU(3)_c$ gauge sector), thereby putting it on the same ground as
that of EW particle physics ($SU(2)\times U(1)$). If an effective
field theory were constructed in a way fully consistent with QCD,
then it should describe nuclear processes as correctly and as
accurately as dictated by QCD if one worked hard enough and
computed all the things that are required within the given
framework. We have seen how this works out in the case of
$\pi$-$\pi$ scattering and $\pi$-$N$ scattering at low energies.
Here one encounters neither puzzles nor any indication that
something is going basically wrong that requires a new insight. I
believe that the same will be the case with EFT's in nuclear
physics when things are developed well enough. It's just a matter
of hard work. Needless to say, we won't be able to calculate
everything that way in nuclear physics: Our computational power is
simply too limited. However for processes that are accessible to a
systematic EFT treatment, the procedure should work out all right.

To me, the more important ra\^ \i son d'\^etre of EFT in nuclear
physics is two-fold: (1) to be able to make precise
error-controlled calculations of certain processes important for
fundamental issues of physics that cannot be provided by the
standard nuclear physics approach (which I will call ``SNPA"); (2)
to make predictions, whether qualitative or quantitative, for
processes that the SNPA cannot access, such as phase transitions
under extreme conditions. If the EFT that we develop is to
reproduce the fundamental theory and if the fundamental theory
represents Nature, then the ultimate goal should be to make
predictions that can in some sense be trusted and verified
eventually. At present this aspect is not as widely recognized as
it merits to be by the physics community.

In this series of lectures, I will develop the notion that when
phrased in the EFT languages, the ``old physics" encapsulated in
SNPA and the ``new physics" contained in QCD may be continuously
connected. This notion will be developed in terms of a circle of
``dualities" implicit in the phenomenon called ``Cheshire Cat
Principle." I will try to develop the arguments that quarks in QCD
are connected to baryons in hadronic spectrum and likewise gluons
to mesons in the sense that a Cheshire-Cat phenomenon is
operative.

\section{Lecture I: The Cheshire Cat Principle}\label{I}
\setcounter{equation}{0} 
\renewcommand{\theequation}{\mbox{\ref{I}.\arabic{equation}}}
\itt In this first lecture, I address the question: Are the
low-energy properties of hadrons sensitive to the size of the
region in which quarks and gluons of QCD are ``confined"? The
Cheshire Cat Principle (CCP) states that they should not be. The
important point to underline here is that we are talking about how
the degrees of freedom associated with the quarks and gluons are
connected to the degrees of freedom associated with physical
hadrons. The CCP addresses this problem in terms of the spaces
occupied by the respective degrees of freedom [NRZ,RHO:PR]. It
simply says that the naive picture of the ``confinement" as
understood in the old days of the MIT bag model is sometimes not
correct. This point will appear later in a different context.
\subsection{The Cheshire Cat Model in (1+1) Dimensions}
\itt Here, I will use a model to develop the picture, rather than
a theory, of EFT. The aim is to describe the general theme that at
low energies, the ``confinement size" has no physical meaning in
QCD. The thing we get out is that in terms of the space time, the
region occupied by the microscopic variables of QCD, i.e., quarks
and gluons, can be made big or small or shrunk to zero without
affecting physics. Since doing this in four dimensions is quite
difficult, I will do this in (1+1) dimension.

In two dimensions, the Cheshire Cat principle (CCP) can be
formulated exactly. In the spirit of a chiral bag, consider a
massless free single-flavored fermion $\psi$ confined in a region
(``inside") of ``volume" $V$ coupled on the surface $\del V$ to a
free boson $\phi$ living in a region of ``volume" $\tilde{V}$
(``outside"). We can think of the ``inside" the line segment $x<R$
and the ``outside" the line segment $x>R$ where $R$ is the
``surface." Of course in one-space dimension, the ``volume" is
just a segment of a line but we will use this symbol in analogy to
higher dimensions. We will assume that the action is invariant
under global chiral rotations and parity. Interactions invariant
under these symmetries can be introduced without changing the
physics, so the simple system that we consider captures all the
essential points of the construction. The action contains three
terms\footnote{Our convention is as follows. The metric is
$g_{\mu\nu}={\rm diag} (1,-1)$ with Lorentz indices $\mu,\nu=0,1$
and the $\gamma$ matrices are in Weyl representation,
$\gamma_0=\gamma^0=\sigma_1$, $\gamma_1=-\gamma^1=-i\sigma_2$,
$\gamma_5=\gamma^5=\sigma_3$ with the usual Pauli matrices
$\sigma_i$.}

\be S=S_V + S_{\tilde{V}} +S_{\del V} \label{action} \ee where
\be
S_V &=& \int_{V} d^2x \bar{\psi}i\gamma^\mu\del_\mu \psi +\cdots,\label{Sin}\\
S_{\tilde{V}} &=& \int_{\tilde{V}} d^2x \frac{1}{2} (\del_\mu
\phi)^2 +\cdots \label{Sout} \ee and $S_{\del V}$ is the boundary
term which we will specify shortly. Here the ellipsis stands for
other terms such as interactions, fermion masses etc. consistent
with the assumed symmetries of the system on which we will comment
later. For instance, there can be a coupling to a $U(1)$ gauge
field in (\ref{Sin}) and a boson mass term in (\ref{Sout}) as
would be needed in Schwinger model. Without loss of generality, we
will simply ignore what is in the ellipsis unless we absolutely
need it. Now the boundary term is essential in making the
connection between the two regions. Its structure depends upon the
physics ingredients we choose to incorporate.

Before going further, one caveat here. We will call $\phi$ the
``pion." Now in (1+1) dimensions, continuous symmetries do not
spontaneously break, so there are no Goldstone bosons. One way to
avoid this caveat is to think that one dimenional space is
actually the radial coordinate of a sphere in 3-dimensional space.
It is in this sense we will be talking about chiral symmetry.

We will assume that chiral symmetry holds on the boundary even if
it is, as in Nature, broken both inside and outside by mass terms.
As long as the symmetry breaking is gentle, this should be a good
approximation since the surface term is a $\delta$ function. We
should also assume that the boundary term does not break the
discrete symmetries ${\cal P}$, ${\cal C}$ and ${\cal T}$. Finally
we demand that it give no decoupled boundary conditions, that is
to say, boundary conditions that involve only $\psi$ or $\phi$
fields. These three conditions are sufficient in (1+1) dimensions
to give a unique term
 \be S_{\del V}=\int_{\del
V} d\Sigma^\mu \left\{\frac{1}{2}n_\mu \bar{\psi}
e^{i\gamma_5\phi/f} \psi\right\} \ee with the $\phi$ ``decay
constant" $f=1/\sqrt{4\pi}$ where $d\Sigma^\mu$ is an area element
with $n^\mu$ the normal vector, {\ie}, $n^2=-1$ and picked
outward-normal. As we will mention later, we cannot establish the
same unique relation in (3+1) dimensions but we will continue
using this simple structure even when there is no rigorous
justification in higher dimensions.

At classical level, the action (\ref{action}) gives rise to the
bag ``confinement," namely that inside the bag the fermion (which
we shall call ``quark" from now on) obeys \be
i\gamma^\mu\del_\mu\psi =0 \ee while the boson (which we will call
``pion") satisfies
 \be \del^2 \phi =0 \ee
subject to the boundary conditions on $\del V$
 \be
in^\mu \gamma_\mu \psi &=&- e^{i\gamma_5 \phi/f}\psi, \label{conf}\\
n^\mu \del_\mu \phi &=& f^{-1}\bar{\psi}(\frac{1}{2}
n^\mu\gamma_\mu \gamma_5) \psi.\label{conf1}
 \ee Equation
(\ref{conf}) is the familiar ``MIT confinement condition" which is
simply a statement of the classical conservation of the vector
current $\del_\mu j^\mu=0$ or
$\bar{\psi}\frac{1}{2}n^\mu\gamma_\mu \psi=0$ at the surface while
Eq. (\ref{conf1}) is just the statement of the conserved axial
vector current $\del_\mu j_5^\mu=0$ (ignoring the possible
explicit mass of the quark and the pion at the
surface)\footnote{Our definition of the currents is as follows:
$j^\mu =\bar{\psi}\frac{1}{2}\gamma^\mu\psi$,
$j_5^\mu=\bar{\psi}\frac{1}{2}\gamma^\mu\gamma_5\psi$.}. The
crucial point of our argument is that these classical observations
are invalidated by quantum mechanical effects. In particular while
the axial current continues to be conserved, the vector current
fails to do so due to quantum anomaly.

There are several ways of seeing that something is amiss with the
classical conservation law, with slightly different
interpretations. The easiest way is as follows. Imagine that the
quark is ``confined" in the space $-\infty \leq r \leq R$ with a
boundary at $r=R$. Now the vector current $j_\mu
=\bar{\psi}\gamma_\mu \psi$ is conserved inside the ``bag"
 \ben
\del^\mu j_\mu=0, \ \ \ \ \ r<R. \een
 If one integrates this from
$-\infty$ to $R$ in $r$, one gets the time-rate change of the
fermion ({\ie}, quark) charge
 \ben \dot{Q}\equiv \frac{d}{dt}
Q=2\int_{-\infty}^R dr\del_0 j_0=2\int_{-\infty} ^R dr \del_1
j_1=2j_1 (R)\label{vanomaly}
 \een
which is just
 \ben \bar{\psi}n^\mu \gamma_\mu \psi, \ \ \ \ r=R.
 \een
This vanishes classically as we mentioned above. But this is not
correct quantum mechanically because it is not well-defined
locally in time. In other words, $\psi^\dagger (t) \psi
(t+\epsilon)$ goes like $\epsilon^{-1}$ and so is singular as
$\epsilon\rightarrow 0$. To circumvent this difficulty which is
related to vacuum fluctuation, we regulate the bilinear product by
point-splitting in time \ben j_1 (t)=\bar{\psi}
(t)\frac{1}{2}\gamma_1 \psi (t+\epsilon), \ \ \ \ \
\epsilon\rightarrow 0. \een Now using the boundary condition \ben
i\gamma_1 \psi (t+\epsilon)&=& e^{i\gamma_5 \phi (t+\epsilon)/f}
\psi (t+\epsilon)\nonumber\\
&\approx& e^{i\gamma_5 \phi (t)/f} [1+i\epsilon \gamma_5
\dot{\phi} (t)/f] e^{\frac{1}{2}[\phi(t),\phi(t+\epsilon)]} \psi
(t+\epsilon),
 \ \ \ \ r=R
\een and the commutation relation \ben [\phi (t), \phi
(t+\epsilon)]=i\ {\rm sign}\ \epsilon, \een we obtain \ben j_1
(t)=-\frac{i}{4f}\epsilon\dot{\phi} (t) \psi^\dagger (t)\psi
(t+\epsilon) =\frac{1}{4\pi f} \dot{\phi} (t) + O(\epsilon), \ \ \
\ r=R \label{flow} \een where we have used $\psi^\dagger (t)\psi
(t+\epsilon)=\frac{i}{\pi\epsilon}+ [{\rm regular}]$. Thus quarks
can flow out or in if the pion field varies in time. But by fiat,
we declared that there be no quarks outside, so the question is
what happens to the quarks when they leak out? They cannot simply
disappear into nowhere if we impose that the fermion (quark)
number is conserved. To understand what happens, rewrite
(\ref{flow}) using the surface tangent \ben
t^\mu=\epsilon^{\mu\nu} n_\nu. \een We have \ben t\cdot\del
\phi=\frac{1}{2f}\bar{\psi} n\cdot \gamma\psi
=\frac{1}{2f}\bar{\psi}t\cdot \gamma\gamma_5 \psi, \ \ \ \ r=R
\label{bosoniz2} \een where we have used the relation
$\bar{\psi}\gamma_\mu\gamma_5\psi=
\epsilon_{\mu\nu}\bar{\psi}\gamma^\nu \psi$ valid in two
dimensions. Equation (\ref{bosoniz2}) together with (\ref{conf1})
is nothing but the bosonization relation \ben \del_\mu \phi=
f^{-1}\bar{\psi} (\frac{1}{2}\gamma_\mu\gamma_5)\psi
\label{bosonization} \een at the point $r=R$ and time $t$. As is
well known, this is a striking feature of (1+1) dimensional fields
that makes one-to-one correspondence between fermions and bosons.

Equation (\ref{vanomaly}) with (\ref{flow}) is the vector anomaly,
{\ie}, quantum anomaly in the vector current: The vector current
is not conserved due to quantum effects. What it says is that the
vector charge which in this case is equivalent to the fermion
(quark) number inside the bag is not conserved. Physically what
happens is that the amount of fermion number $\Delta Q$
corresponding to $\Delta t\dot{\phi}/\pi f$ is pushed into the
Dirac sea at the bag boundary and so is lost from inside and gets
accumulated at the pion side of the bag wall. This accumulated
baryon charge must be carried by something residing in the meson
sector. Since there is nothing but ``pions" outside, it must be be
the pion field that carries the leaked charge. This means that the
pion field supports a soliton. This is rather simple to verify in
(1+1) dimensions. This can also be shown to be the case in (3+1)
dimensions. In the present model, we find that one unit of fermion
charge $Q=1$ is partitioned as \ben
Q&=& 1=Q_V + Q_{\tilde{V}},\\
Q_{\tilde{V}} &=& \theta/\pi,\nonumber\\
Q_V &=& 1-\theta/\pi\nonumber \een with \ben \theta=\phi (R)/f.
\een We thus learn that the quark charge is partitioned into the
bag and outside of the bag, without however any dependence of the
total on the size or location of the bag boundary. In
(1+1)-dimensional case, one can calculate other physical
quantities such as the energy, response functions and more
generally, partition functions and show that the physics does not
depend upon the presence of the bag. We could work with quarks
alone, or pions alone or any mixture of the two. If one works with
the quarks alone, we have to do a proper quantum treatment to
obtain something which one can obtain in mean-field order with the
pions alone. In some situations, the hybrid description is more
economical than the pure ones. {\it The complete independence of
the physics on the bag -- size or location-- is called the
``Cheshire Cat Principle" (CCP) with the corresponding mechanism
referred to as ``Cheshire Cat Mechanism" (CCM).} Of course the CCP
holds exactly in (1+1) dimensions because of the exact
bosonization rule. There is no exact CCP in higher dimensions
since fermion theories cannot be bosonized exactly in higher
dimensions but a fairly strong case of CCP can be made for (3+1)
dimensional models. Topological quantities like the fermion (quark
or baryon) charge satisfy an exact CCP in (3+1) dimensions while
nontopological observables such as masses, static properties and
also some nonstatic properties satisfy it approximately but rather
well.

So far we have been treating the quark as ``colorless", in other
words, in the absence of a gauge field. Let us consider that the
quark carries a $U(1)$ charge $e$, coupling to a $U(1)$ gauge
field $A^\mu$. We will still continue working with a
single-flavored quark, treating the multi-flavored case later.
Then inside the bag, we have essentially the Schwinger model,
namely, (1+1)-dimensional QED. It is well established that the
charge is confined in the Schwinger model, so there are no charged
particles in the spectrum. If now our ``leaking" quark carries the
color (electric) charge, this will at first sight pose a problem
since the anomaly obtained above says that there will be a leakage
of the charge by the rate \ben
\dot{Q^c}=\frac{e}{2\pi}\dot{\phi}/f  . \een This means that the
charge accumulated on the surface will be \ben \Delta
Q^c=\frac{e}{2\pi}\phi (t,R)/f. \een Unless this is compensated,
we will have a violation of charge conservation, or breaking of
gauge invariance. This is unacceptable. Therefore we are forced to
introduce a boundary condition that compensates the induced
charge, {\ie}, by adding a boundary term \ben \delta S_{\del
V}=-\int_{\del V} d\Sigma \frac{e}{2\pi}\epsilon^{\mu\nu} n_\nu
A_\mu \frac{\phi}{f} \een with $n^\mu=g^{\mu\nu}n_\nu$. The action
is now of the form  \ben S=S_V+S_{\tilde{V}}+S_{\del
V}\label{saction} \een with \ben S_V&=& \int_V d^2x
\left\{\bar{\psi} (x)\left[i\del_\mu-eA_\mu\right]\gamma^\mu
\psi (x) -\frac{1}{4}F_{\mu\nu}F^{\mu\nu}\right\},\\
S_{\tilde{V}}&=& \int_{\tilde{V}} d^2x \left\{\frac{1}{2}(\del_\mu
\phi (x))^2
-\frac{1}{2}m_{\phi}^2\phi (x)^2\right\},\\
S_{\del V}&=& \int_{\del V} d\Sigma \left\{\frac{1}{2}\bar{\psi}
e^{i\gamma^5\phi/f}\psi -\frac{e}{2\pi}\epsilon^{\mu\nu} n_\mu
A_\nu\frac{\phi} {f} \right\}. \een We have included the mass of
the $\phi$ field, the reason of which will become clear shortly.

In this example, we are having the ``pion" field (more precisely
the soliton component of it) carry the color charge. In general,
though, the field that carries the color could be different from
the field that carries the soliton. In fact, in (3+1) dimensions,
it is the $\eta^\prime$ field that will be coupled to the gauge
field (although $\eta^\prime$ is colorless) while it is the pion
field that supports a soliton. But in (1+1) dimensions, the
soliton is lodged in the $U(1)$ flavor sector (whereas in (3+1)
dimensions, it is in $SU(n_f)$ with $n_f\geq 2$). For simplicity,
we will continue our discussion with this Lagrangian.
\subsection{The Mass of the Scalar $\phi$}
\itt We now illustrate how one can calculate the mass of the
$\phi$ field using the action (\ref{saction}) and the CCM
following [NRWZ]. This exercise will help understand a similar
calculation of the mass of the $\eta^\prime$ in (3+1) dimensions.
The additional ingredient needed for this exercise is the $U_A
(1)$ (Adler-Bell-Jackiw) anomaly which in (1+1) dimensions takes
the form~\footnote{In (3+1) dimensions, it has the form \ben
\del_\mu j_5^\mu=-\frac{e^2}{32\pi^2} \epsilon_{\mu\nu\rho\sigma}
F^{\mu\nu} F^{\rho\sigma}.\nonumber \een}
  \ben \del_\mu
j_5^\mu=-\frac{e}{4\pi}\epsilon_{\mu\nu}F^{\mu\nu}
=-\frac{1}{2\pi} eE.\label{abj1}
 \een
Instead of the configuration with the quarks confined in $r<R$,
consider a small bag ``inserted" into the static field
configuration of the scalar field $\phi$ which we will now call
$\etap$ in anticipation of the (3+1)-dimensional case we will
consider later. Let the bag be put in $-\frac{R}{2}\leq r \leq
\frac{R}{2}$. The CCP states that the physics should not depend on
the size $R$, hence we can take it to be as small as we wish. If
it is small, then the charge accumulated at the two boundaries
will be $\pm \frac{e}{2\pi}\phi/f$. This means that the electric
field generated inside the bag is \ben E=F_{01}=\frac{e}{2\pi}
(\phi/f). \een

Therefore from the $U_A (1)$ anomaly (\ref{abj1}), the axial
charge created or destroyed (depending on the sign of the scalar
field) in the static bag is \ben 2\int_V dr \del_\mu
j_5^\mu=2(-j_1^5|_{\frac{R}{2}} + j_1^5|_{-\frac{R}{2}})
=-\frac{e^2}{2\pi^2} (\phi/f)R. \een {}From the bosonization
condition (\ref{bosonization}), we have \ben
2f\left(\del_1\phi|_{\frac{R}{2}}-\del_1\phi|_{-\frac{R}{2}}\right)
=\frac{e^2}{2\pi^2}(\phi/f)R. \een Now taking $R$ to be
infinitesimal by the CCP, we have, on cancelling $R$ from both
sides, \ben \del^2 \phi-\frac{e^2}{4 \pi^2 f^2}\phi=0 \een which
then gives the mass, with $f^{-2}=4\pi$, \ben
m_{\phi}^2=\frac{e^2}{\pi},\label{schmass} \een the well-known
scalar mass in the Schwinger model which can also be obtained by
bosonizing the $(QED)_2$. A completely parallel reasoning has been
used to calculate the $\eta^\prime$ mass in (3+1) dimensions
[NRWZ, NRZ].

\begin{figure}[hbt]
\vskip 0.cm
 \centerline{\epsfig{file=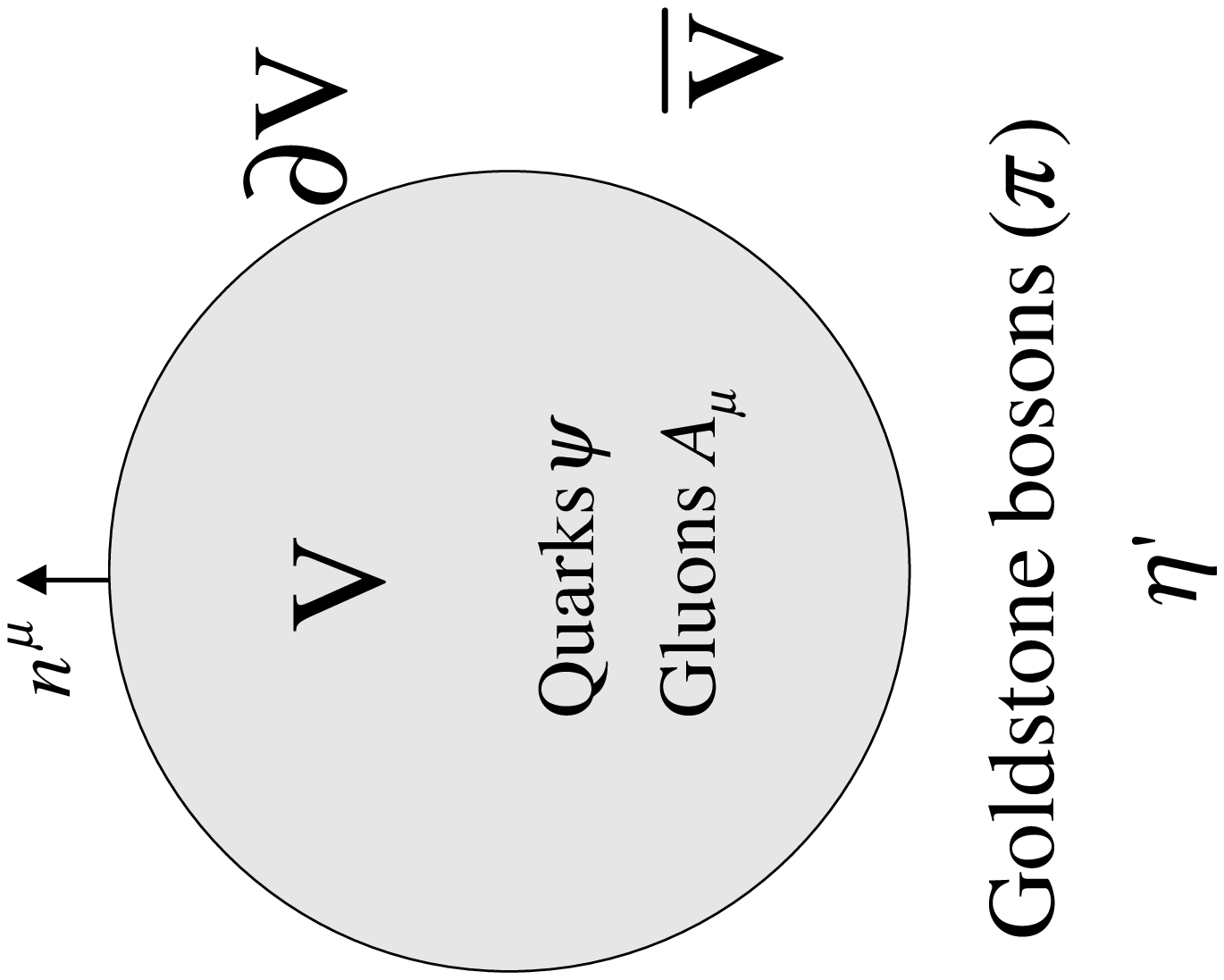,width=9cm,angle=-90}}
\vskip -2.cm \caption{\small A spherical chiral bag for
``deriving" a four-dimensional Cheshire Cat model. $\psi$
represents the doublet quark field of u and d quarks for $SU(2)$
flavor or the triplet of u, d and s for $SU(3)$ flavor and $\pi$
the triplet $\pi^+, \pi^-, \pi^0$ for $SU(2)$ and the octet
pseudoscalars $\pi$, $K$, $\bar{K}$ and $\eta$ for $SU(3)$. The
axial singlet meson $\eta^\prime$ figures in axial
anomaly.}\label{bag}
\end{figure}
\subsection{The Cheshire Cat Model in (3+1) Dimensions}
\itt Arguments paralleling the (1+1) dimensional model leads to
the Lagrangian appropriate for (3+1) dimensions. We will just
write it down and define the meanings of the terms involved. See
[NRWZ, NRZ]. A cartoon of the model is given by Fig. \ref{bag}.
 \ben
S&=& S_V+S_{\tilde{V}}+S_{\del V},\label{cheshire}\\
S_V&=& \int_V d^4x \left(\bar{\psi}i\not\!\!{D}\psi -\frac{1}{2}
{\rm tr}\ F_{\mu\nu}F^{\mu\nu}\right)+\cdots\nonumber\\
S_{\tilde{V}}&=&\frac{f^2}{4}\int_{\tilde{V}} d^4x \left(\Tr\
\del_\mu U^\dagger \del^\mu U +\frac{1}{4N_f}
m^2_{\eta^\prime}({\rm ln}U-{\rm ln}U^\dagger)^2
\right) +\cdots  + S_{WZW},\nonumber\\
S_{\del V}&=& \frac{1}{2}\int_{\del V} d\Sigma^\mu\left\{(n_\mu
\bar{\psi} U^{\gamma_5}\psi) +i\frac{g^2}{16\pi^2}{K_5}_\mu (\Tr\
{\rm ln} U^\dagger-\Tr\ {\rm ln} U)\right\}.
 \een
Here $f$ is the pion decay constant $f\equiv f_\pi\sim 93$ MeV,
$U$ is the unitary chiral field
 \ben
U=e^{i\eta^\prime/f_0}e^{2i\pi/f},
\ U^{\gamma_5}=e^{i\gamma_5\eta^\prime/f_0}e^{2i\gamma_5\pi/f}\\
f_0\equiv \sqrt{N_f/2} f,
 \een
$S_{WZW}$ is the Wess-Zumino-Witten term defined in the space-time
volume $\tilde{V}$ and $K_{5\mu}$ is the Chern-Simons current
 \ben K_5^\mu=\epsilon^{\mu\nu\alpha\beta}
(A_\nu^a F_{\alpha\beta}^a-\frac{2}{3} gf^{abc} A_\nu^a A_\alpha^b
A_\beta^c).
 \een

If in accordance with the CCP one were to shrink the bag to a
point, that is, let $R\rightarrow 0$, then one would be left over
with only the meson fields. In terms of the $U$ field alone, this
would be an infinite series in derivatives, the leading order
terms taking the form
 \be
S&=&\frac{f^2}{4}\int_{V+\tilde{V}} d^4x \Tr\ \del_\mu U^\dagger
\del^\mu U +\cdots + S_{WZW}\label{largeN}
 \ee
with the WZW term defined in the {\it whole space}
$(V+\tilde{V})$. The ellipsis stands for lots of higher order
(higher derivative, mass etc.) terms which need to be included
depending upon the processes treated. What comes out here is just
the generalization of Skyrme-type theory, representing QCD at
large $N_c$. One may introduce heavier meson fields replacing part
of the ellipsis. In Lectures III and IV, the vector mesons $\rho$,
$\omega$ will figure explicitly and play an important role. The
part of the WZW term belonging to the volume $V$ arises through an
interplay between anomalies inside $V$ and the surface terms as
the volume $V$ is shrunk to a point and precisely makes up the
complement to the WZW term of the volume $\tilde{V}$. How this
comes about -- which is quite intricate -- is described in [NRZ].

The Lagrangian (\ref{cheshire}) has all the relevant symmetries
and anomalies that should give a correct representation of QCD for
large $N_c$. No one has yet fully studied the contents of this
theory as there are certain aspects of the equations which have
not been fully worked out. So far there is nothing which indicates
that the CCP fails in Nature.

I don't have the time for discussing it here but let me just
mention that in the form (\ref{largeN}), the effective Lagrangian
should describe not only the baryons but also finite nuclei,
nuclear matter and dense matter. The current development involves
mainly classical solutions of the skyrmion-type theory for the
systems with a finite number of baryons, i.e., ``finite nuclei,"
which are interesting from a mathematical point of view but do not
resemble the real nuclei. It remains to be seen what happens when
the solutions are quantized. An extremely interesting recent
development is to study infinite matter within the Skyrme
Lagrangian. Again this is at the classical level, with the
discovery of a variety of crystal structures at ``high density" (
see [PMRV, SCHWINDT] for a recent effort) but again it remains to
be quantized. Approaching dense matter through the skyrmion matter
is relevant to the study of phase structure of hadronic matter
under compression needed for unravelling the structure of compact
stars.

\subsection{The ``Proton Spin" Problem}
\itt The so-called ``proton spin" problem had defied the CCP until
recently and it is only now [LMPRV] that it is resolved within the
framework of the CC model (\ref{cheshire}). This problem has been
studied extensively not only in the context of hadronic models but
also from the point of view of QCD proper. In fact there are many,
most likely equivalent, ways, some more in line with QCD than
others, but we will not go into this matter. The objective of the
present discussion is not so much to ``explain" the resolution of
the problem but to illustrate the subtlety involved in the way the
CC manifests in this problem.

 Since the ``proton spin" issue is connected with the flavor
singlet axial matrix element $a_0$, we have to define what the
current is in the chiral bag model (CBM) (\ref{cheshire}).  We
write the current in the CBM as a sum of two terms, one from the
interior of the bag and the other from the outside populated by
the meson field $\eta^\prime$ (how to account for the Goldstone
pion fields is known; the mechanism is identical to what was
described above in the context of the (1+1) dimensional picture
and it will be taken into account for the baryon charge leakage)
 \be A^\mu =A^\mu_B \Theta_B + A^\mu_M
\Theta_M.\label{current}
 \ee
For notational simplicity, we will omit the flavor index in the
current. We shall use the short-hand notations $\Theta_B=\theta
(R-r)$ and $\Theta_M=\theta (r-R)$ with $R$ the radius of the bag.
We interpret the $U_A (1)$ anomaly as given in this model by
 \be
\partial_\mu A^\mu =
\frac{\alpha_s N_F}{2\pi}\sum_a \vec{E}^a \cdot \vec{B}^a
\Theta_{B}+ f m_\eta^2 \eta \Theta_{M}.\label{abj}
 \ee
We are assuming here that in the nonperturbative sector outside of
the bag, the only relevant $U_A (1)$ degree of freedom is the
massive $\eta^\prime$ field. This allows us to write
 \be
A^\mu_M = A^\mu_\eta = f\del^\mu \eta
 \ee
with the divergence
 \be \del_\mu A^\mu_\eta = fm_\eta^2 \eta. \ee
It turns out to be more convenient to write the current as
 \be
A^\mu=A_{B_{Q}}^\mu + A_{B_{G}}^\mu + A_\eta^\mu\label{sep}
 \ee
such that
 \be
\partial_\mu (A_{B_{Q}}^\mu + A_\eta^\mu) &=& f m_\eta^2 \eta
\Theta_{M},\label{Dbag}\\
\partial_\mu A_{B_G}^\mu &=&
\frac{\alpha_s N_F}{2\pi}\sum_a \vec{E}^a \cdot \vec{B}^a
\Theta_{B}. \label{Dmeson}
 \ee
The subindices Q and G imply that these currents are written in
terms of quark and gluon fields respectively. In writing
(\ref{Dbag}), the up and down quark masses are ignored. Since we
are dealing with an interacting theory, there is no unique way to
separate the different contributions from the gluon, quark and
$\eta$ components. In particular, the separation we adopt,
(\ref{Dbag}) and (\ref{Dmeson}), is neither unique nor gauge
invariant  although the sum is without ambiguity. This separation,
however, is found to lead to a natural partition of the
contributions in the framework of the bag description for the
confinement mechanism that is used here.

We now discuss individual contributions from each term.

$\bullet$ {\bf The quark current $A_{B_Q}^\mu$}

The quark current is given by
 \be
 A^\mu_{B_Q} = \bar{\Psi} \gamma^\mu \gamma_5 \Psi
 \ee
where $\Psi$ should be understood to be the {\it bagged} quark
field. Therefore the quark current contribution to the FSAC is
given by
 \be a^0_{B_Q} = \langle p|\int_B d^3r \bar{\Psi} \gamma_3
\gamma_5 \Psi|p\rangle. \label{aq}
 \ee The calculation of this type of
matrix elements in the CBM is nontrivial due to the baryon charge
leakage between the interior and the exterior through the Dirac
sea. But we know how to do this in an unambiguous way. A complete
account of such calculations and references can be found in the
literature [RHO:PR, NRZ]. The leakage produces an $R$ dependence
which would otherwise be absent. One finds that there is no
contribution for zero radius, that is in the pure skyrmion
scenario for the proton. The contribution grows as a function of
$R$ towards the pure MIT result although it may never reach it
even at infinite radius.

$\bullet$ {\bf The meson current $A^\mu_\eta$}

\begin{figure}
\centerline{\epsfig{file=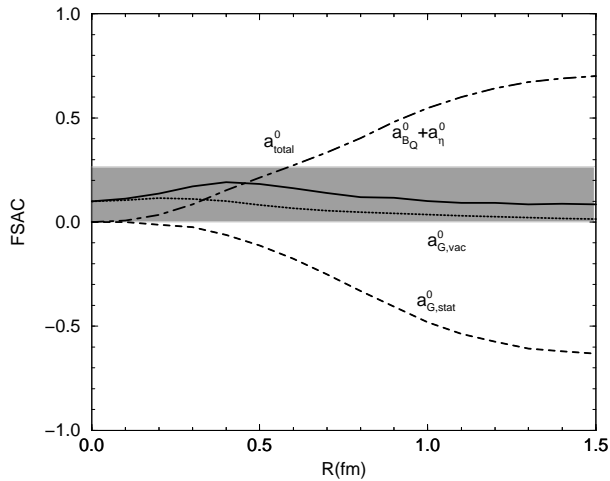, width=9cm}} \caption{\small
Various contributions to the flavor singlet axial current of the
proton as a function of bag radius and comparison with the
experiment: (a) quark plus $\eta$ (or ``matter") contribution
($a^0_{B_Q} + a^0_\eta$), (b) the contribution of the static
gluons due to quark source ($a^0_{G,stat}$), (c) the gluon Casimir
contribution ($a^0_{G,vac}$), and (d) their sum ($a^0_{total})$.
The shaded area corresponds to the range admitted by
experiments.}\label{cc}
\end{figure}

Due to the coupling of the quark and $\eta$ fields at the surface,
we can simply write the $\eta$ contribution in terms of the quark
contribution,
 \be a^0_\eta = \frac{1 +y_\eta}{2(1+y_\eta)
+y_\eta^2} \langle p|\int_B d^3r \bar{\Psi} \gamma_3 \gamma_5
\Psi|p\rangle. \label{aeta}
 \ee
where $y_\eta = m_\eta R$.  Since the $\eta$ field has no
topological structure, its contribution also vanishes in the
skyrmion limit.  Due to baryon charge leakage, however, this
contribution increases slowly as the bag increases. This
illustrates how the dynamics of the exterior can be mapped to that
of the interior by boundary conditions. We may summarize the
analysis of these two contributions by stating that no trace of
the CCP is apparent from the ``matter" contribution. As shown in
Fig. \ref{cc}, there is a sensitive dependence on $R$. Thus if the
CCP were to emerge, the only possibility would be that the gluons
do the miracle!

Let us turn to the gluon contribution. The gluon current is split
into two pieces
 \be
A^\mu_{B_G} = A^\mu_{G,stat}  + A^\mu_{G,vac}.
 \ee
The first term arises from the quark and $\eta$ sources, while the
latter is associated with the properties of the vacuum of the
model. One might worry that this contribution could not be split
in these two terms without double counting. However this worry is
unfounded. Technically, it is easy to check it by noticing that
the former acts on the quark Fock space and the latter on the
gluon vacuum. Thus, one can interpret the former as a one gluon
exchange correction to the quantity. One can also show this
intuitively by making the analogy to the condensate expansion in
QCD, where the perturbative terms and the vacuum condensates enter
additively to the lowest order.

$\bullet$ {\bf The gluon static current $A^\mu_{G,stat}$}

Let us first describe the static term.

To the leading order, we can the $\eta$ coupling. Afterwards, the
$\eta$ contribution can be added. The boundary conditions for the
gluon field would correspond to the original MIT ones. The quark
current is the source term that remains in the equations of motion
after performing a perturbative expansion in the QCD coupling
constant, i.e., the quark color current \be g{\bar \Psi}_0
\gamma_\mu \lambda^a \Psi_0 \ee where the $\Psi_0$ fields
represent the lowest cavity modes. In this lowest mode
approximation, the color electric and magnetic fields are given by
 \be \vec{E}^a = g_s
\frac{\lambda^a}{4\pi} \frac{\hat{r}}{r^2} \rho (r) \label{ef} \ee
\be \vec{B}^a = g_s \frac{\lambda^a}{4\pi}\left( \frac{\mu
(r)}{r^3}(3 \hat{r} \vec{\sigma} \cdot \hat{r} - \vec{\sigma}) +
(\frac{\mu (R)}{R^3} + 2 M(r)) \vec{\sigma}\right) \label{bf}
 \ee
where $\rho$ is related to the quark density $\rho^\prime$
as\footnote{Note that the quark density that figures here is
associated with the color charge, {\it not} with the quark number
(or rather the baryon charge) that leaks due to the hedgehog
pion.}
 \be \rho (r,\Gamma)=\int_\Gamma^r ds \rho^\prime
(s)\label{density}\nonumber
 \ee and $\mu, M$ to the vector current
density
 \be \mu (r) &=& \int_0^r ds \mu^\prime (s),\nonumber\\ M
(r)&=& \int_r^R ds \frac{\mu^\prime (s)}{s^3}.\nonumber
 \ee The
lower limit $\Gamma$ is taken to be zero in the MIT bag model --
in which case the boundary condition is satisfied only {\it
globally}, that is, after averaging -- and $\Gamma=R$ in the so
called {\it monopole solution} -- in which case, the boundary
condition is satisfied {\it locally}. We take the latter since
consistency with the CCP condition rules out the MIT condition.

We now proceed to introduce the $\eta$ field. We perform the same
calculation with however the color boundary conditions coming from
(\ref{cheshire}) -- which are modified by the color anomaly from
the MIT ones -- taken into account. In the approximation of
keeping the lowest non-trivial terms, the boundary conditions
become
 \be \hat{r}\cdot \vec{E}_{stat}^a=\frac{N_F
g^2}{8\pi^2 f} \hat{r}\cdot \vec{B}^a_g \eta (R)\label{Eg} \ee
 \be
\hat{r}\times \vec{B}_{stat}^a=-\frac{N_F g^2}{8\pi^2 f}
\hat{r}\times \vec{E}^a_g \eta(R).\label{Bg}
 \ee
Here $\vec{E}^a_g$ and $\vec{B}^a_g$ are the lowest order fields
given by (\ref{ef}) and (\ref{bf}) and $\eta (R)$ is the meson
field at the boundary. The $\eta$ field is given by
 \be \eta
(\vec{r}) = -\frac{g_{NN\eta}}{4\pi M} \vec{S} \cdot \hat{r}
 \frac{1+m_\eta r}{r^2} e^{-m_\eta r}
\label{eta}\ee where the coupling constant is determined from the
surface conditions.

Note that the magnetic field is not affected by the new boundary
conditions, since $\vec{E}^a_g$ points into the radial direction.
The effect on the electric field is just a change in the charge,
i.e.,
 \be \rho_{stat}(r) = \rho(r,\Gamma) + \rho_\eta(R) \ee where
\be \rho_\eta (R) = \frac{N_F g^2}{64 \pi^3 M}
\frac{g_{NN\eta}}{f} (1+y_\eta) e^{-y_\eta}. \ee

The contribution to the FSAC arising from these fields is
determined from the expectation value of the anomaly
 \be
a^0_{G,stat} = \langle p|-\frac{N_F\alpha_s}{\pi} \int_B d^3r x_3
\vec{E}^a_{stat} \cdot \vec{B}^a_{stat}|p\rangle . \label{astat}
 \ee One finds that including the $\eta$ contribution in
$\rho_{stat} (r)$ brings a non-negligible modification to the FSAC
but does not modify the result qualitatively. The result as one
can see in Fig. \ref{cc} shows that this contribution is zero at
$R=0$ but increases as $R$ increases but with the sign opposite to
that of the matter field, largely cancelling the $R$ dependence of
the matter contribution. We should remark here that there is a
drastic difference between the effect of the MIT-like electric
field and that of the monopole-like electric field: The former is
totally incompatible with the Cheshire Cat property whereas the
latter remains consistent independently of whether or not the
$\eta$ contribution is included in $\rho_{stat}$.

$\bullet$ {\bf The gluon Casimir current $A^\mu_{G,vac}$}

Up to this point, the FSAC is zero for $R=0$ and non-zero for
$R\neq 0$. This is in principle a violation of the CCP although
the magnitude of the violation may be small. We now show that it
is the vacuum contribution through Casimir effects that the CCP is
restored. The calculation is subtle involving renormalization of
the Casimir effects, the details of which are to be found in the
paper by Lee et al. [LMPRV]. Here we summarize the salient feature
of the contribution.

The quantity to calculate is the gluon vacuum contribution to the
flavor singlet axial current of the proton, which can be done by
evaluating the expectation value
 \begin{equation}
\langle 0_B| -\frac{N_F\alpha_s}{\pi} \int_V d^3r x_3 ( \vec{E}^a
\cdot \vec{ B}^a) |0_B\rangle \label{f1}
 \end{equation}
where $|0_B\rangle$ denotes the vacuum in the bag. To calculate
this, we invoke at this point the CCP which states that at low
energy, hadronic phenomena do not discriminate between QCD degrees
of freedom (quarks and gluons) on the one hand and meson degrees
of freedom (pions, etas,...) on the other, provided that all
necessary quantum effects (e.g., quantum anomalies) are properly
taken into account. If we consider the limit where the $\eta$
excitation is a long wavelength oscillation of zero frequency, the
CCP asserts that it does not matter whether we choose to describe
the $\eta$, in the interior of the infinitesimal bag, in terms of
quarks and gluons or in terms of mesonic degrees of freedom. This
statement, together with the color boundary conditions, leads to
an extremely simple and useful {\it local} formula,
 \begin{equation}
\vec{ E}^a \cdot \vec{ B}^a \approx -\frac{N_F g^2}{8\pi^2}
\frac{\eta}{f} \frac12 G^2, \label{f2}
 \end{equation}
where only the term up to the first order in $\eta$ is retained in
the right-hand side. Here we adapt this formula to the CBM. This
means that the couplings are to be understood as the average bag
couplings and the gluon fields are to be expressed in the cavity
vacuum through a mode expansion.  That the surface boundary
condition can be interpreted as a local operator is a rather
strong CCP assumption which while justifiable for small bag
radius, can only be validated \`a posteriori by the consistency of
the result. This procedure is the substitute to the condensates in
the conventional discussion.

Substituting Eq.(\ref{f2}) into Eq.(\ref{f1}) we obtain
 \begin{eqnarray}&& \hskip -2em
\langle 0_B| -\frac{N_F\alpha_s}{\pi} \int_V d^3r x_3 (\vec{ E}^a
\cdot \vec{ B}^a) |0_B\rangle \nonumber\\
&& \approx  \left(-\frac{N_F\alpha_s}{\pi} \right)
\left(-\frac{N_F g^2}{8\pi^2} \right) \frac{y(R)}{f_0}
\langle p|S_3|p\rangle (N_c^2-1) \nonumber\\
&& \hskip 3em \sum_n \int_V d^3r (\vec{ B}_n^* \cdot \vec{ B}_n -
\vec{ E}_n^* \cdot \vec{ E}_n) x_3 \hat{x}_3 , \label{ps1}
 \end{eqnarray}
where we have used that $\eta$ has a structure of $(\vec{
S}\cdot\hat{ r}) y(R)$. Since we are interested only in the first
order perturbation, the field operator can be expanded by using
MIT bag eigenmodes (the zeroth order solution). Thus, the
summation runs over all the classical MIT bag eigenmodes. The
factor $(N_c^2-1)$ comes from the sum over the abelianized gluons.

The next steps are the  numerical calculations to evaluate the
mode sum appearing in Eq.(\ref{ps1}): (i) introduction of the heat
kernel regularization factor to classify the divergences appearing
in the sum and (ii) subtraction of the ultraviolet divergences.
These procedures -- which involve an intricate manipulation -- are
described in [LMPRV]. The result is shown in Fig. \ref{cc}. Though
the magnitude is small compared with the others, it is important
at small $R$ to restore, within the CBM scheme, the CCP. The net
result which is small due to the intricate cancellation between
the matter contribution and the gluon contribution compares well
with the experimental range quoted in the literature.

The lesson from this calculation is that neither the matter
contribution nor the gluon contribution to $a^0$, both of which
are gauge-non-invariant and CCP-violating, is physical. Only the
total which is gauge invariant is physical and CCP-preserving.

\section{Lecture II: Effective Field Theories for Dilute Matter and
Superdense Matter}\label{II}
\setcounter{equation}{0} 
\renewcommand{\theequation}{\mbox{\ref{II}.\arabic{equation}}}
\subsection{Strategy of EFT}
\itt I will start by briefly stating the principal idea of EFT
relevant for low-energy processes that we will be considering.
This is a nutshell presentation.

Picking a scale given by the cutoff $\Lambda$, one first divides a
generic field $\Phi$ -- that consists of $all$ degrees of freedom,
bosonic as well fermionic -- into the ``high" field $\Phi_H$ lying
above $\Lambda$ and the ``low" field $\Phi_L$ lying below
$\Lambda$, i.e., $\Phi=\Phi_H +\Phi_L$ and integrate out from the
generating functional or partition function $Z$ the ``high"
component (in which we are not specifically interested) and write
$Z$ in terms of $\Phi_L$ only. Defining the action in Euclidean
space as
 \be
S[\Phi]=S_0[\Phi_L]+S_0[\Phi_H]+S_{I}[\Phi_L,\Phi_H]
 \ee
where $S_0$ stands for the non-interacting part of the action and
$S_{I}$ for the interaction part, one writes
 \be
Z=\int [d\Phi_L]\e^{-S^{eff} [\Phi_L]}
 \ee
where
 \be
e^{-S^{eff} [\Phi_L]}=e^{-S_0 [\Phi_L]}\int [d\Phi_H] e^{-S_0
[\Phi_H]} e^{-S_{I} [\Phi_L,\Phi_H]}.
 \ee
This defines the mode elimination referred to as ``decimation."
The next step is to write $S^{eff}$ in terms of integrals over
local fields
 \be
S^{eff} [\Phi_L]=\int^\Lambda \sum_i C_i (\Lambda) Q_i\label{seff}
 \ee
where $Q_i$ are polynomials of the local $\Phi_L$ fields. It
should be emphasized that in general, when certain fields are
integrated out, the resulting effective action is not always
local, so in some cases (as discussed later), the localization can
be a bad approximation. For the moment we will proceed assuming
that the localization can be done.

There are typically two ways that the expansion (\ref{seff}) can
be effectuated. If one has a theory that is well defined above and
below $\Lambda$, i.e., over the whole space (like QED taken as a
fundamental theory, not in a context of an effective theory that
unifies all the interactions), then one can compute the
coefficients $C_i$ as a function of $\Lambda$ from the theory. QCD
that concerns us is not like this. Although QCD is a full theory
by itself, the two regimes, above and below $\Lambda$, are not
simply describable in terms of the same degrees of freedom. For
instance, in the approach we will take, we imagine the degrees of
freedom above $\Lambda$ to be given in terms of the microscopic
variables, quarks and gluons, and those below $\Lambda$ by
hadrons. In this case, we cannot simply compute the coefficients
$C_i$  from theory (except perhaps on lattice) but we have to
resort to experiments. In doing this, one has to inject a certain
dose of intuition and guessing. The major task in doing this is to
assess and minimize the errors committed in truncating the series.

The next thing to do is to do the scale counting in writing down
the series (\ref{seff}). Given an action in D dimensions
 \be
S^{eff}=\int d^Dx {\calL}^{eff}
 \ee
with the effective Lagrangian expressed in terms of the naive
dimension of the field operators as
 \be
{\calL}^{eff}=\calL_{\leq D} +\calL_{D+1} +\calL_{D+2} +\cdots
 \ee
one identifies the term $\calL_{\leq D}$ to be ``naively
renormalizable" and all the rest ``naively non-renormaiizable."
They have the following scaling property. To be specific, consider
the scalar field theory
 \be
S=\int d^Dx\left\{\frac 12 (\del_\mu \phi)^2 -\frac 12 m^2 \phi^2
+a\phi^4 +b\phi^6 +\cdots\right\}.
 \ee
We want to determine how each operator in the action scales when
one scales the space-time $x$,
 \be
x\rightarrow s x, \ \ \ s>1.\label{scaling}
 \ee
To do this, we have to define the standard measure. We do this by
decreeing that the kinetic term in the action remains unscaled
under (\ref{scaling}). Now since $\del\rightarrow s^{-1}\del$, we
have
 \be
\int d^Dx(\del_\mu\phi)^2\rightarrow s^{D-2+2[\phi]}\int
d^Dx(\del_\mu\phi)^2.
 \ee
We want this term unchanged, so for $D=4$, we find that
$[\phi]=-1$. Such a term that remains unchanged under scaling is
referred to as ``marginal."

Now the rest follows immediately. The $\Phi^4$ term is also
marginal. The mass term scales as $s^{D-2}=s^2$.  Terms scaling as
$s^{n}$ with $n>0$ are referred to as ``relevant." The $\phi^6$
term goes as $s^{-2}$ etc. Terms scaling as $s^m$ with $m<0$ are
called ``irrelevant." {\it What these terms mean physically is
that as the probe momentum/energy goes down, that is, as
$s\rightarrow \infty$, marginal terms remain unchanged, irrelevant
terms die away and relevant terms blow up."} Something unusual
happens when the relevant terms take over and that something is
related to phase transitions. Interesting physics can be lodged in
the irrelevant terms although they get suppressed at low energy.
In nuclear physics, as we will see later (e.g., the solar $pp$ and
$hep$ processes), what nuclear physicists have been calling
``short-range correlations" can be identified in the irrelevant
terms in the action.

The next step is then to sum the series (\ref{seff}) including
loop graphs of the same power counting. In doing this, the
coefficients $C_i$ enter as counter terms to do the
renormalization.

Suppose now we have the series. The given generating functional
(or partition function) $Z$ then gives physical amplitudes
${\calA}$. Since where one puts the cutoff is arbitrary, physical
amplitudes should not depend upon what specific $\Lambda$ one
chooses, assuming of course one knows how to pick all necessary
degrees of freedom. This gives rise to the renormalization-group
(RG) invariance,
 \be
d\calA/d\Lambda=0
 \ee
which leads to a relation for the coefficients $C_i$
 \be
dC_i (\Lambda)/d\Lambda=\calF_i(\Lambda)\label{rge}
 \ee
where $\calF_i$ are given functions of $\Lambda$. These are the
renormalization group equations, known as Wilson equations or
Callan-Symanzik equations depending upon what kind of theory one
is dealing with. Our formulation given below is closer to the
Wilsonian approach.

I have been describing the procedure in too general a term to this
point. Let me be somewhat more specific so the discussion given
later on the RG flows of hidden local symmetry (HLS) theory can be
understandable. Suppose that $\Phi_L$ consists of a set of, say,
three fields labelled $\phi_A, \phi_B, \phi_C$ defined at a scale
$\Lambda_a$, i.e., $(\phi_A, \phi_B, \phi_C)\in \Phi_L$. Now
imagine that we are decimating the degrees of freedom from
$\Lambda_a$ down to $\Lambda_b<\Lambda_a$ at which point the field
$\phi_A$ decouples. All three fields $\phi_{A,B,C}$ contribute to
the flows in this range. Next as one decimates from $\Lambda_b$ to
$\Lambda_c<\Lambda_b$ at which point the field $\phi_B$ decouples,
only $\phi_B$ and $\phi_C$ will contribute to the flow. From
$\Lambda_c$ down, then only $\phi_c$ will contribute to the flow
etc. Thus while one is lowering the cutoff in terms of
energy/momentym scale, one is also ``integrating out" associated
degrees of freedom. In the next lecture, we will see that the
various RG scales are intricately tied to the external
disturbances such as density and/or temperature (i.e.,
density/temperature dependence of the RG scales) and one has to
keep track of how the scales vary as a function of
density/temperature in following the flows.

It turns out in certain cases that the right-hand side of
(\ref{rge}) vanishes. In this case, we have a set of  ``fixed
points." The variety of fixed points, e.g., conformal fixed point,
vector manifestation (VM) fixed point, Fermi-liquid fixed point
etc., will play a primordial role in our developments.
\subsection{EFT for Two-Nucleon Systems}
\itt As the first case, let us consider two-nucleon systems at
very low energy. These systems are very well understood in the
SNPA and as such, we will learn nothing new from the discussion
given in this subsection as far physics as is concerned. However
we do gain some insight as to how EFT works in the regime where
things are well understood.

For low energy nuclear physics, the relevant degrees of freedom
are the local fields of proton, neutron and pions. We imagine that
all other degrees of freedom have been integrated out, with their
imprints left in the higher-dimension terms. Thus the field spaces
are
 \be
(n, p, \pi)&\in& \Phi_L,\nonumber\\
(\rho, \omega, a_1, \Delta, \ {\rm glueballs}, \cdots, J/\psi,
\cdots) &\in& \Phi_H.
 \ee In this case, the series $C_iQ_i$ in
(\ref{seff}) takes the form
 \be
 f_\pi^2\Lambda^2(\pi/f_\pi)^m (\del/\Lambda)^n
 (N/f_\pi\sqrt{\Lambda})^r
 \ee
where $f_\pi$ is the pion decay constant and $\Lambda$ here is the
chiral cutoff $\sim 4\pi f_\pi \sim 1$ GeV. Chiral perturbation
theory for pion-nucleon interactions is developed with this series
in terms of the power ${\cal N}=m+n+r$.

If we further restrict ourselves to energy or momentum much less
than the pion mass scale, we can even integrate out the pions as
well. See [BEANE]. This is not a very good idea, however, if one
wants to study response functions of two-nucleon (as well as
multi-nucleon) systems (due to what is known as ``chiral filter
mechanism" in nuclei) as we will do in the next subsection where
we will keep the pion fields explicitly. But for the discussion at
hand, namely, S-wave scattering of two nucleons, eliminating the
pions is justified. Treating the nucleon as nonrelativistic, the
effective Lagrangian can be written as
 \be
\calL^{eff}=N^\dagger\left(i\del_t+\frac{\vec{\nabla}^2}{2M}\right)
-\frac 12 [C^s (N^\dagger N)^2+C^t (N^\dagger\vec{\sigma} N)^2]
+\cdots\label{lowL}
 \ee
where $M$ is the nucleon mass, the $C$'s are unknown constants to
be fixed from experiments, and the ellipsis stands for higher
nucleon fields and/or higher derivative terms. We shall ignore
these higher dimension terms for the discussion.

\begin{figure}[hbt]
\vskip -0.cm
 \centerline{\epsfig{file=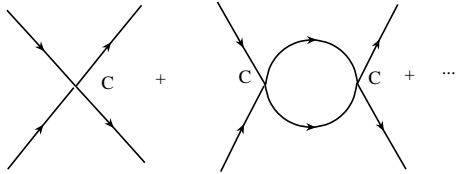,width=7cm,angle=0}}
\vskip -1.5cm \caption{\small Feynman graphs for nucleon-nucleon
scattering via the four-nucleon contact interaction Lagrangian.
}\label{figtw1}
\end{figure}

Given the Lagrangian (\ref{lowL}), one can do the standard
calculation in terms of a Schr\"odinger equation provided the
four-Fermi contact term is suitably regularized. One can also
calculate an infinite set of Feynman diagrams and sum them. The
latter is totally equivalent to the former (see [PKMR:CO]). It is
however more transparent to do the latter.  Summing the Feynman
graphs of Fig. \ref{figtw1} to all orders, one gets the amplitude
 \be
\calA=-\frac{C}{1-C(GG)}
 \ee
where $(GG)$ stands for the two-nucleon propagators which in the
CM frame is
 \be
(GG)=\int \frac{d^3q}{(2\pi)^3}
\frac{1}{E-q^2/M+i\epsilon}\label{gg}
 \ee
with $\sqrt{ME}=|\vec{k}|\equiv k$.
The integral (\ref{gg}) diverges, so needs to be regularized. We
cut the integral at $\Lambda$ -- to be specified below -- and
obtain~\footnote{If one blindly uses dimensional regularization
(DR), the linear divergence is absent. Something wrong with this
result since in effective field theories, the power divergences
that are ``killed" by the DR are physical quantities and should be
taken into account. This problem arises also in hidden local
symmetry theory discussed below [HY:PR] when one wants to describe
phase transitions in HLS framework. What one has to do is to
subtract divergences present at a dimension less than four --
dimension 3 in this case and dimension 2 in HLS. This is called
``power divergence subtraction" in DR. No such rigamaroles are
needed in the cutoff regularization. Of course one has to be
careful in using the cutoff regularization if there is chiral
symmetry.}
 \be
(GG)= -\frac{M}{4\pi} \left(\frac{\Lambda}{\pi} +ik\right).
 \ee
Thus
 \be
\calA=-\frac{C}{1+\frac{CM}{4\pi}
(\frac{\Lambda}{\pi}+ik)}.\label{A1}
 \ee
In general the coefficient $C$ will depend on where the cutoff
$\Lambda$ is put. For a given $\Lambda$ it may be obtained from
lattice. Unfortunately, we have no data on this. So we will resort
to experiments to fix it. To do this, we write the amplitude in
terms of the S-wave phase shift $\delta$,
 \be
\calA=\frac{4\pi}{M} \frac{1}{k\cot\delta -ik}\approx
-\frac{4\pi}{M} \frac{1}{\frac 1a +ik}\label{A2}
 \ee
where the second approximate equality comes from the effective
range formula,  $k \cot\delta=-1/a +\frac 12 r_0 k^2 +\calO
(k^4)$. Comparing (\ref{A1}) and (\ref{A2}), we get
 \be
C(\Lambda)=\frac{4\pi}{M} \left(\frac{1}{-\frac{\Lambda}{\pi}
+1/a}\right).\label{Clambda}
 \ee
Note that the amplitude (\ref{A1}) with (\ref{Clambda}) satisfies
the RG invariance condition $\Lambda
\frac{d\calA^{-1}}{d\Lambda}=0$ as it should.

What is interesting with this formula is that in Nature, the
scattering length is huge~\footnote{I.e., $a_0\equiv
a(^1S_0)=-23.7$ fm and $a_1\equiv a(^3S_1)=5.4$ fm.}compared with
the typical hadronic scale $\sim 0.2$ fm and also with the cutoff
scale which is expected to be of order $\sim (3 m_\pi)^{-1}$ so
the $1/a$ term in (\ref{Clambda}) plays an unimportant role. Let
us assume that the scattering length is infinite. Then we see that
[MEHEN, BEANE]
 \be
\frac{d[\Lambda C(\Lambda)]}{d\Lambda}=0.
 \ee
This means that there is a fixed point, called ``scale invariant
fixed point."~\footnote{Actually it turns out to be a conformally
invariant fixed point [MEHEN].} To see that the theory for
$a=\infty$ is scale-invariant, one notes that the four-Fermi
interaction Lagrangian becomes
$\calL_{4fermi}=-\frac{2\pi}{M\Lambda} (N^\dagger N)^2$, so the
action remains invariant under the scale change
$\Lambda\rightarrow s^{-1}\Lambda$ and $N\rightarrow s^{-3/2} N$.

What's the big deal with the scale invariance of the theory? It
means that the S-wave nucleon-nucleon interaction at low energy is
dominated by the scale-invariant fixed point. So it makes a good
sense to fluctuate around this fixed point to do two-nucleon
physics at low energy. But was it essential for describing
two-nucleon interactions to know that there is such a fixed point?
The answer is no. Nuclear physicists who knew nothing about it
have been doing precision nuclear physics all along and got the
right answers that can be compared with experiments.

I should note that as I did here with the cutoff regularization,
there is nothing so special about the large scattering length. It
is only in doing naive dimensional regularization that all sorts
of bizarre things are encountered. Clearly were the $\Lambda$ term
in (\ref{Clambda}) absent as in naive DR, the coefficient $C$
would be ``unnaturally"  big and the expansion would make no
sense. This led some people to think that an EFT in nuclear
physics needed new ingredient. But it is the naive regularization
that is at fault, not the EFT: there is nothing that indicates
that EFT is sick, in whatever form the EFT was formulated. This
point is discussed in [PKMR:CO].

What about the pions? The pions are found to be somewhat awkward
in keeping the counting kosher and some people attempted to treat
the pions as perturbative. This may be OK for some low-energy
scattering but treating the pions perturbatively not only spoils
chiral invariance but also makes certain problems needlessly
harder if not impossible. We know that the pion is essential for
the deuteron structure. For one thing, the torus structure implied
by the pion tensor force -- which is visible experimentally, e.g.,
through electron scattering -- cannot possibly be brought in by
perturbation. Furthermore the ``chiral filter mechanism" mentioned
below [KDR] works marvellously well whenever soft-pion mechanisms
are operative and guide us how to do a model-independent
calculation even when they are not operative. The pion enters
indispensably in this story which is the topic of the next
subsection.
\subsection{Predictive EFT}
\itt As I mentioned before, one of the objectives of doing EFT is
to confirm that one can do a systematic calculation that is
consistent with QCD of nuclear properties. One can for instance
work out nuclear forces -- nucleon-nucleon as well as
multi-nucleon -- to as high an order in the EFT counting as
possible. This may be possible for two-nucleon interactions and
perhaps eventually three-nucleon interactions. But physics-wise,
it seems to me that {\it the best one can hope to achieve here is
to arrive at the accuracy obtained by those accurate
phenomenological potentials available in the literature}. It is
possible and perhaps instructive to gauge the consistency of the
methods used by the standard nuclear physics approach (SNPA) but I
do not see what new physics can be learned from such hard work. It
will of course be gratifying to verify that ``nuclear physicists
knew what they were doing and that they were doing it correctly."
The bottom line is that the harder one works here, the better it
will come close to the phenomenological results. The most
up-to-date effort in this direction can be found in [EPEL1,EPEL2].

I believe that it is in making $predictions$ -- and not struggling
with parameter-full exercises -- that cannot be made by the SNPA
in which the real power of EFT lies. After all, that is the
ultimate objective of a fundamental theory, a feat that cannot be
expected of models.

\subsubsection{Chiral filter mechanism}
\itt I will follow Weinberg's original chiral perturbation scheme
in which the pion is put on the same footing as contact
non-derivative four-nucleon interactions with the pion mass
incorporated by means of perturbative unitarity. There is a bit of
problem with the power counting when the pion is present {\it ab
initio} and this has been discussed extensively by several people
[BEANE]. I would not like to get into that matter as I believe it
is a technical matter that does not seem to be important in most
of the processes considered so far.

I will take it for granted that a sophisticated phenomenology with
light nuclei can supply us accurate wave functions with which one
can calculate response functions. Although $n$-body potentials for
$n>2$ cannot be unambiguously determined in this way, two-body and
possibly three-body potentials could be determined with great
accuracy.  And there is a growing evidence that this is the case.
Solving the Schr\"odinger equation with such potentials
corresponds to summing to all orders a subset of ``reducible
graphs" in the EFT expansion, with the ``irreducible graphs"
subsumed to be taken into account in the ``accurate potentials" up
to some high order. In exploiting these accurate wave functions
that emerge from such calculations in the context of an EFT, it
would of course be great to have a clear idea what is included and
what is not included in the potentials used. In some cases, one of
which is discussed here, this is possible. Now given such wave
functions, can one calculate response functions measured in
precision experiments as accurately as possible with the
possibility of controlling the theoretical errors one commits ?
This question can be answered affirmatively for the calculation of
responses to slowly-varying EW fields.

\begin{figure}[htb]
\vskip -0.0cm
 \centerline{\epsfig{file=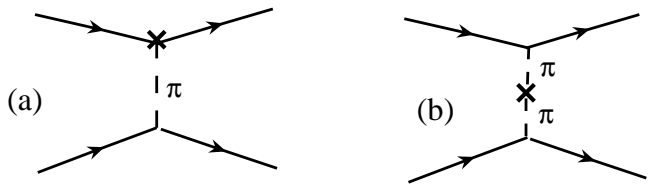,width=12cm,angle=0}}
\vskip -3.5cm \caption{\small Two-body currents with one soft-pion
exchange which  dominate whenever unsuppressed by kinematics or
symmetry. The cross stands for the current. Both (a) and (b)
contribute to the vector current but only (a) contributes to the
axial current. }\label{softpi}
\end{figure}

Consider the matrix element of the vector current $J_\mu^a$ and
the axial-vector current $J_{5\mu}^a$. We are interested in
calculating $\la f|J_\mu|i\ra$ where $i$ and $f$ denote
respectively the initial and final nuclear states. This current
effective in $A$-body system can be decomposed into
 \be
J_\mu=\sum_{n=1}^A J_\mu^{(n)}.
 \ee
Customarily, except for unusual cases, an example of which we will
enocunter below, one-body terms are leading in the power counting
(chiral counting in chiral-invariant theories) -- and they $are$
numerically, so the theorist's task is to compute the matrix
elements of higher-body currents. Since the EW current will act
only once for very weak and slowly-varying interactions, it
suffices to systematically count the chiral orders of the {\it
irreducible graphs} contributing to the current. This program was
initiated many years ago [CR:71]. In terms of exchanges of mesons
between the nucleons in interaction, the dominant ``correction" to
the leading one-body is the 2-body contribution with the exchange
of one {\it soft-pion} in Fig. \ref{softpi}.
The chiraL filter mechanism states [KDR] that {\it whenever the
one soft-pion exchange is allowed, unsuppressed by kinematics and
symmetry, the soft-pion exchange two-body current dominates the
correction with higher (chiral order) terms suppressed typically
by an order of magnitude and calculable reliably}. This means that
one can compute the transition matrix elements with high accuracy
within the framework of chiral perturbation theory. Conversely if
the one-soft-pion term is suppressed, then corrections to the
leading term are several orders higher in the power counting or
shorter-ranged and cannot be accessed reliably with only a finite
number of terms. In this case, ordinary chiral perturbation theory
is not much of power and one has to resort to a different strategy
than ordinary chiral perturbation theory. A simple analysis of the
graphs of Fig. \ref{softpi} shows that the space component of the
vector current and the time component of the axial current are
protected by the chiral filter mechanism and the remainders are
not [KDR]. One beautiful example that supports this observation is
the thermal $np$ capture process [PMR] which involves the space
component of the vector current, that is, isovector $M1$ operator,
 \be
n+p\rightarrow d+\gamma,
 \ee
the cross section of which is predicted (in the sense that there
are no free parameters) with a theoretical error of $\lsim 1\%$
with the prediction agreeing perfectly with the experiment. The
other example is the axial charge transition in heavy nuclei
 \be
A (0^\pm)\rightarrow A^\prime (0^\mp)\ e\ \nu \ \ \ \Delta T=1.
 \ee
It has been confirmed in forbidden $\beta$ transitions that the
enhancement due to the soft-pion exchange graph with a suitable
scaling due to density in the chiral Lagrangian (described below)
can be $\sim 100\%$ in heavy nuclei such as Pb [KR:91, RHO:MIG,
BR:PR01].
\subsubsection{Predictions for the solar $pp$ and $hep$ processes}
\itt It comes as a surprise that even when the chiral filter
mechanism is suppressed and hence the soft-pion contribution is
absent, under certain circumstances (to be described below), one
can still make accurate predictions without being obstructed by
unknown parameters. I discuss two cases here which are quite
important for astrophysics.~\footnote{As far as I can see, the
purist approach that eschews the good old wave functions but
strictly adheres to the order-by-order power counting cannot
possibly obtain a genuine prediction for the $hep$ process without
unknown parameters. See later for a bet.} The processes I will
consider, recently discussed in [PMSV:pp, PMSV:hep], are
 \be
p+p&\rightarrow& d+e^+ +\nu_e,\label{pp}\\
^3{\rm He} +p&\rightarrow& ^4{\rm He} + e^+ +\nu_e.\label{hep}
 \ee

These two processes are important for the solar neutrino problem
that has bearing on the issues of neutrino mass and stellar
evolution. What we are concerned with here is neither the neutrino
mass nor the stellar evolution but with the strong interaction
input -- that is, accurate nuclear matrix elements of the weak
current --  which is of course essential for the main issues. This
problem turns out to be highly nontrivial, particularly for the
$hep$ process (\ref{hep}), for the following reasons. A naive
(chiral) power counting shows that the dominant contribution
should come from the space part of the axial current, in
particular, the single-particle Gamow-Teller (GT) operator, since
the lepton momentum transfer is small. However the chiral filter
argument says that the soft-pion correction is suppressed for the
space component of the axial current and hence corrections to the
leading one-body current would inevitably involve shorter-distance
physics. In terms of the chiral power counting relative to the
leading single-particle GT operator which formally is of $\calO
(p^0)$, corrections to the GT would start at
next-to-next-to-next-to-leading order (N$^3$LO or $\calO (p^3)$).
(There are also small contributions from the axial charge operator
but this operator is protected by the chiral filter and hence is
accurately calculable.)  If the single-particle GT matrix element
does not have $anomalous$ suppression, then this term should make
up the bulk of the amplitude and the N$^3$LO correction, even with
an inherent uncertainty, would not affect the total significantly.
This is indeed the case for the $pp$ process (\ref{pp}): Here the
single-particle term makes up typically more than 95 \% of the
decay rate. However this is not the case for (\ref{hep}) because
of an ``accidental" suppression of the single-particle GT term
caused by the spatial symmetry mismatch between the initial and
final wave functions. Furthermore the situation is even more
acerbated  since the leading correction term comes with an
opposite sign to the single-particle term with a comparable
magnitude. Due to the cancellation, the resulting decay rate could
differ by orders of magnitude depending on the theory. Thus for
the process in question, it is imperative that the suppressed
correction term be calculated with high accuracy as emphasized in
the context of EFT in [PKMR:TAI].

The details of the solution to this problem are rather complex but
the general idea of how it comes about is rather simple. I shall
describe this in as simple a way as possible. Since other terms
than the GT are straightforward and unambiguous, let me focus on
the GT.

Since the long-range soft-pionic contribution is made inoperative
in the GT channel, it is the short-range interaction that
surfaces. In the standard nuclear physics approach, this aspect of
physics is interpreted as ``short-range correlation." In the
present framework of EFT, this short-distance N$^3$LO current
receives contributions from both ``hard" pionic part and heavy
mesonic part, zero-ranged as the heavy mesons are integrated out.
Denote the matrix element of the finite-range pionic part by
 \be
 M_\pi=\int d^3 r F_\pi (r)\label{pion}
 \ee
and that of the zero-ranged pionic part and heavy-meson parts by
 \be
M_{HM}=\int d^3r F_{HM} (r).\label{heavy}
 \ee
The coefficients of the operators in (\ref{pion}) as well as the
pionic part of (\ref{heavy}) are of course known. However the rest
of the terms in (\ref{heavy}), e.g., the counter terms, are not
known. Since they must depend on the cutoff imposed, they cannot
be obtained by saturating with a set of known heavy mesons. They
can only be determined from experiments if data are available.

To have a contact with SNPA, we work in coordinate space. The
quantities $F_{\pi,{\rm HM}} (r)$ contain information on the
``exact" wave functions (with the phenomenological potentials
fitted to an ensemble of experimental data). By the procedure, the
wave functions embody not only the physics ingredient that figures
in the calculation of the currents calculated to N$^3$LO but also
much -- though perhaps not all -- of short-distance interactions
that are of higher order than N$^3$LO. If the currents were
calculated to all orders in the chiral counting and the wave
functions corresponded to the same order, then the integrals would
be well-defined without any further regularization. However our
currents are calculated to a given order, i.e., N$^3$LO, and the
wave function to an order presumably higher than N$^3$LO.
Therefore,  the integrals will diverge and to make sense, an
ultraviolet regularization is needed. In SNPA, one customarily
cuts off the integral at a ``hard-core" size $r=r_c$. Such a hard
core kills all zero-range terms (including {\it all counter
terms}) in (\ref{heavy}) as well as cuts short-range piece of the
(known) finite-range terms in (\ref{pion}). I will call this SNPA
procedure ``hard-core regularization (HCR)." If the process is
dominated by long-range interactions as in the case where the
chiral filter mechanism is operative, this prescription is
expected -- and verified -- to be reliable. The prime example is
the thermal $np$ capture mentioned above. However in the present
problem, the principal action comes from the short-distance part,
so it is evident that the HCR prescription will give a result that
strongly depends on the hard-core radius, thus upsetting the tenet
of EFT and hence predictivity.

This is where the basic idea of EFT comes to help. The strategy of
EFT is to regularize the operators in both (\ref{pion}) and
(\ref{heavy}) in such a way that the integrals are well-defined
and the sum of (\ref{pion}) and (\ref{heavy}) comes out
independent of the cutoff one imposes to the order
considered~\footnote{In the papers cited above, this was done to
$\calO (p^3)$ but were this done to $\calO (p^4)$ or higher, the
strategy would be essentially the same. Clearly this cannot be
done at the leading order where the corrections to the GT operator
do not come in.}. This regularization will be referred to as
``modified hard-core regularization (MHCR)." This is not a trivial
feat and there seems to be much misunderstanding on this in the
community of nuclear EFT.

First what is the relevant scale of the cutoff? Since the lightest
degrees of freedom that enter in the short-distance physics
embedded in the counter terms are the scalar $\sigma$, $\rho$,
$\omega$ mesons, we expect the cutoff to be in the vicinity of the
$\sigma$ mass $\gsim 500$ MeV.

Next what is expected of a bona-fide EFT? An EFT requires that the
cutoff dependence in (\ref{pion}) -- that reflects defect in
short-distance physics in (\ref{pion}) -- be cancelled by the
cutoff dependence in (\ref{heavy}). This implies that the
coefficients of the counter terms will be cutoff dependent; the
stronger it will be, the more short-ranged the physics is. Now in
order for this procedure to work, we need an independent
experimental source that determines the cut-off dependent
parameters in (\ref{heavy}). For the processes in question
(\ref{pp}) and (\ref{hep}), it is the tritium beta decay that
supplies the crucial link:
 \be
^3{\rm H}\rightarrow ^3{\rm He} + e^- +\bar{\nu}_e.\label{tritium}
 \ee
What turns out to be remarkable is that the same linear
combination of counter-term operators figures in all three
processes (\ref{pp}), (\ref{hep}) and (\ref{tritium}). The reason
for this ``miracle" is that the same symmetry is operative in this
GT channel. Furthermore since one is dealing with a rather
short-ranged interaction, the same dynamics prevails for the
two-body current whether it takes place in two-body, 3-body or
4-body system. Thus once the single unknown term (called
$\hat{d}^R$ in [PMSV:pp, PMSV:hep]) is fixed for a given cutoff
from (\ref{tritium}), there are no free parameters for the
processes (\ref{pp}) and (\ref{hep}). This makes the calculation
of the delicate correction term firmer since three-body and
four-body currents (suppressed by power counting) turn out to be
totally negligible numerically.

To give an idea what happens, let me give what is called ``S
factor" which carries the nuclear information needed. With the
error estimated from the cutoff dependence in the range $500 \leq
\Lambda/{\rm MeV} \leq 800$, the results are
 \be
S_{pp}&=& 3.94 (1\pm 10^{-3})\times 10^{-25}\ {\rm MeV-barn},\\
S_{hep}&=& (8.5\pm 1.4)\times 10^{-20}\ {\rm MeV-barn}.
 \ee
I would say that these results are one of the most accurate
predictions ever made in nuclear physics. The reason why the $pp$
can be calculated with such a greater precision is that the main
term is the single-particle GT and the correction term which is
quite small $\lsim 1$ \% can be more or less controlled. Because
of the accidental suppression of the single-particle GT, such an
accuracy cannot be attained in the $hep$ process. Nonetheless, the
result is remarkable, considering that in the past the uncertainty
was orders of magnitude.

To conclude, whenever the chiral filter is operative, the dominant
single-particle contribution and the soft-pion correction thereof
are accurately given by the matrix elements computed with the SNPA
wave functions. In this case, it is possible to argue that the
wave functions do make an {\it integral part} of the EFT itself.
The aficionados of the puristic power counting will agree to this
statement since it is possible to do a rigorous calculation fully
consistent with chiral counting at each order and come to the same
result.  But there is nothing really gained in adhering to the
counting rule since one learns nothing beyond what we already know
from SNPA. When the chiral filter mechanism is not operative, the
same aficionados will face difficulty in calculating, say, the
$hep$ process because it involves short distance and hence high
orders, perhaps much higher than what can be managed with the
given experimental data.~\footnote{This is like a centipede who is
unable to make steps by being over-worried about the detailed
motion of each foot.}

I should note that in the approach presented here, the possible
error in counting brought in by the ``exact" wave functions with
$\calO (p^3)$ currents is most likely compensated by the
regularization procedure that assures the cutoff independence.
Exactly the same observation was made in the calculation of the
isoscalar $M1$ and $E2$ matrix elements in the process
$n+p\rightarrow d+\gamma$ [PKMR:M1/E2]. Although these matrix
elements, governed by the chiral-filter-unprotected operators, are
suppressed with respect to the dominant isovector $M1$ matrix
element by $\sim$ three orders of magnitude, the same
regularization scheme used above produced a prediction with a very
small theoretical error bar. What is in action is again a
universal feature associated with the short-distance interaction
not easily accessible by a low-order chiral perturbation
expansion. This prediction can be potentially checked by
experiments.
\subsubsection{Experimental tests}
\itt There are a number of experimental indications that the
predictions or postdictions for those processes protected by the
chiral filter are valid. The situation for the predictions
discussed above where short-distance regularization, i.e., the
modified hard-core regularization (MHCR), enters has not yet been
confirmed. When the parity-violation experiments in the process
$n+p\rightarrow d+\gamma$ are sufficiently refined, one could
perhaps isolate the suppressed isoscalar $M1$ matrix element
($M1S$) and $E2$ matrix element ($E2S$) and check the prediction
based on the MHCR. The presently available data on polarization
observables are not precise enough for the test.

The solar neutrino experiments performed at the SNO and the
super-Kamiokande provide some information on the $hep$ process but
at present, there is little one can say about the value of the
$hep$ amplitude from the observation. Future refined measurements
will perhaps supply the necessary information.

A process somewhat related to the $hep$ process is the $hen$
process
 \be
^3{\rm He}+n\rightarrow ^4{\rm He} +\gamma\label{hen}
 \ee
which involves the same wave function overlap problem. An EFT
analysis using the same strategy used for the $hep$ process might
be illuminating to test some of the ideas used for the latter.
However one should note that as is known since a long time [CR:71]
the structure of multi-body currents is basically different
between the EM vector current that enters in (\ref{hen}) and the
axial-current that enters in (\ref{hep}). It is not obvious that
the short-distance physics that plays a crucial role in
(\ref{hep}) figures in (\ref{hen}) in the same way. This point is
being analyzed and we will have the answer in the near future.

\subsection{EFT for Color-Flavor-Locked (CFL) Dense Matter}
\itt I will now turn to a topic which is perhaps unrelated to what
I discussed above but which can be accessed also by a fundamental
approach anchored on QCD. The EFT for two nucleon systems and the
EFT for superhigh density are the only ones I know that can be
considered ``well founded" from the point of view of QCD.
\subsubsection{Bosonic effective Lagrangian} \itt At
super-high density, QCD is readily tractable, since it becomes
weak coupling so that one can make a clear theoretical statement.
I discuss this problem in preparation for the argument that will
be developed later for the density regime of interest. It is
believed that at large density, a variety of interesting phenomena
such as, e.g. color superconductivity, kaon condensation etc. can
take place and may play an important role in the physics of
compact stars [RW]. The superdense matter that I am dealing with
here is probably not much relevant to Nature that is observable
but it is a matter for which the theory can be clear-cut and
highly instructive. I discuss it here not so much for possible
physical relevance but for developing a general framework for
treating density regimes that are relevant to Nature.

At very high density such that the chemical potential $\mu$ is
much greater than any of the light-quark mass $\mu\gg m_q$, we may
consider the chiral limit of $SU(3)$ flavor symmetry. Although not
rigorously proven, it is believed that there is an effective
attraction in QCD interactions in the color anti-triplet channel
$\bar{\boldmath 3}$ that triggers a Cooper paring of diquarks,
thus producing superconductivity in color. We will assume that the
diquark condensation occurs in such a way that the color and
flavor get locked
 \be
\epsilon_{ab}\la \psi_{iL}^{a\alpha} (\vec{p})\psi_{jL}^{b\beta}
(-\vec{p})\ra=-\epsilon_{ab}\la \psi_{iR}^{a\alpha}
(\vec{p})\psi_{jR}^{b\beta} (-\vec{p})\ra=\Delta
(p^2)\epsilon^{\alpha\beta A}\epsilon_{ijA}\label{condense}
 \ea
where $(\alpha,\beta)$ are color indices, $(i,j)$ flavor indices,
$(a,b)$ spinor indices, $L(R)$ stands for left (right) field and
$\Delta$ is a constant representing the gap. This is the form one
gets considering only the color anti-triplet channel. There can in
principle be condensation in the color sextet ($\boldmath 6$)
which would add a term but we assume, following the workers in the
field, that that term can be ignored. Now since
$\epsilon^{\alpha\beta A}\epsilon_{ijA} = \delta^\alpha_a
\delta^\beta_b - \delta^\alpha_b\delta^\beta_a$, we see that the
color and flavor get locked [RW]. What this means in terms of the
symmetries of the system is as follows. Ignoring $U(1)$ symmetries
associated with the baryon number $U(1)_V$ and the axial $U(1)_A$,
the unbroken global symmetry involved is $G=SU(3)_C\times
SU(3)_L\times SU(3)_R$. Now according to (\ref{condense}), the
color locks to L as well as to R as
 \be
SU(3)_C\times SU(3)_L&\rightarrow& SU(3)_{C+L},\label{C+L}\\
SU(3)_C\times SU(3)_R&\rightarrow& SU(3)_{C+R}.\label{C+R}
 \ee
Now since the color is vectorial, this means that the $L$ and $R$
are broken down to the diagonal $V= L+R$ as well, so effectively
the symmetry breaking is
 \be
G\rightarrow H
 \ee
where
 \be
G&=&SU(3)_C\times SU(3)_L\times SU(3)_R,\\
H&=&SU(3)_{P}
 \ee
with $P=C+V$. Since the chiral symmetry is broken, we have an
octet of pseudoscalar Goldstone bosons which I will denote by
$\pi=\lambda^a\pi^a/2$ with $\lambda$ being the Gell-Mann
matrices. Since the global color symmetry is completely broken,
there are an octet of scalar Goldstone bosons which we will denote
by $s=\lambda^a s^a/2$. I would like to express the coordinates
$\xi\in G/H$ in terms of these two sets of Goldstone fields:
 \be
\xi_{L(R)i}^\alpha=\epsilon_{\alpha\beta\gamma} \epsilon^{ijk}\la
\psi^\beta_{L(R)j}\psi^\gamma_{L(R)k}\ra\label{diquarkxi}
 \ea
where I have dropped spinor indices for economy in notation. Let
$h\in SU(3)_C$ and $g_{L(R)}\in SU(3)_{L(R)}$ be the generators of
the respective transformations. Then the $\xi$ field transforms as
 \ba
\xi_{L(R)}\rightarrow h\xi_{L(R)}g^\dagger_{L(R)}.
 \ea
A convenient parameterization is
 \ba
\xi_L (x)=e^{-i\pi (x)/F_\pi} e^{is(x)/F_\sigma}, \ \
\xi_R(x)=e^{i \pi (x)/F_\pi} e^{is(x)/F_\sigma}.\label{xi}
 \ea
We still have the local gauge invariance associated with the
gluons $G_\mu=\lambda^a G^a_\mu$  -- which will eventually be
spontaneously broken, so the degrees of freedom that we have are
$\pi$, $s$ and $G_\mu$. I will consider the fermions
(quarks/baryons) later. To construct gauge-invariant theory with
these fields, define the covariant derivative
 \ba
D_\mu \xi_{L,R} (x)=(\del_\mu-iG_\mu)\xi_{L,R} (x)
 \ea
and write
 \ba
\hat{\alpha}_{V\mu} (x)&\equiv& (D_\mu \xi_L\cdot\xi_L^\dagger
+D_\mu \xi_R\cdot\xi_R^\dagger) (-i/2),\nonumber\\
\hat{\alpha}_{A\mu} (x)&\equiv& (D_\mu \xi_L\cdot\xi_L^\dagger
-D_\mu \xi_R\cdot\xi_R^\dagger) (-i/2).
 \ea
The leading-order Lagrangian  with the lowest power of derivatives
is
 \ba
\calL_{p^2}=F^2_\pi\Tr(\hat{\alpha}_{A\mu})^2+a F^2_\pi\Tr
(\hat{\alpha}_{V\mu})^2\label{leading}
 \ea
where $a$ is a constant to be determined later. Adding the kinetic
energy term for the gluons, we have
 \ba
\calL^{eff}_{CFL}=\calL_{p^2} -\frac{1}{2g^2}\Tr
(G_{\mu\nu})^2\label{effCFL}
 \ea
with $G_{\mu\nu}=\del_\mu G_\nu -\del_\nu G_\mu -i[G_\mu, G_\nu]$.
The ``gluon" kinetic energy term which would formally be of $\calO
(p^4)$ for massive gluons can actually be of $\calO (p^2)$ under
certain conditions that will be considered below. Of course we
should have other terms of the same order as well as quark mass
terms. For the moment, we continue without them.

The fact that the color-flavor locking is operative is described
by that $F_\pi\neq 0$ and $a\neq 0$. Since the local gauge
symmetry is spontaneously broken, the gluons get Higgsed and
become massive. This can be seen in the unitary gauge which
corresponds to setting $s=0$ in (\ref{xi}). The mass formula is
then
 \be
m_G^2=a F_\pi^2 g^2.
 \ee
I point out that this formula is a familiar one from low-energy
chiral dynamics where it is known as ``KSRF relation". We will
encounter this later in ``hidden local symmetry (HLS)" theory I
will discuss in low-density regime.

The theory we have written down is a low-excitation theory based
on symmetry. In medium, one has to take into account the fact that
Lorentz invariance is broken, so the time and space components of
various quantities have to be distinguished. This can be done
readily, so I won't crowd the equations. Unlike in the low-energy
low-density case,  we have here QCD at hand to work with. Since we
have a full theory valid within the regime, one can integrate out
uninteresting high-energy degrees freedom such as anti-quarks from
the QCD Lagrangian and obtain an effective QCD theory valid for
large chemical potential (see [HONG]) given in terms of the QCD
variables only and hence containing no free parameters. To
distinguish such a Lagrangian from effective Lagrangians given in
terms of hadronic variables referred to as EFT, I will reserve
``effective QCD (EQCD)" for it. The EQCD allows one to calculate
the parameters of the low-energy EFT (\ref{effCFL}) by matching
the latter to it. How this is done in practice can be found in the
references given later. The result is that [SON]
 \be
a_t=a_s=1, \ \ F^2_{\pi t}=\frac{\mu^2 (21-8\ln2)}{36\pi^2}, \ \
F^2_{\pi s}=\frac 13 F^2_{\pi t}
 \ee
where the subscript $t$ ($s$) stands for the time (space)
component. Now for $\mu\rightarrow \infty$, $g(\mu) \sim
(\ln\mu)^{-1}$, so the gluons become super-massive and hence
decouple. In this case, the effective Lagrangian (\ref{effCFL})
becomes
 \be
\calL^{eff}= \frac{F_\pi^2}{4}\Tr (\del_\mu U \del^\mu U^\dagger)
 + {\rm mass\  term}+\cdots\label{CA}
  \ee
 where $U=\xi^\dagger_L\xi_R=e^{2i \pi (x)/F_\pi}$. This is just
the same chiral Lagrangian we have seen before resulting from the
Cheshire Cat model in the limit that the bag is shrunk to zero,
i.e., (\ref{largeN}).

What we have done in arriving at (\ref{CA}) is, exploiting the
``gauge equivalence," to go from the linear $G_{global}\times
H_{local}$ theory to the nonlinear $G/H$ theory. Later in
Bando-Kugo-Yamawaki HLS theory, we will go in the opposite
direction. It is important to realize that here the local gauge
symmetry is
 ``explicit" although ``dressed" in the sense that the group is
 $SU(3)_{P=V+C}$. It is not a ``hidden local symmetry" in the sense
 of hidden gauge symmetry discussed below. We will see that the
 same reasoning holds in the low-density regime for which the HLS
 theory is relevant. This will indicate that explicit local
 symmetry and hidden local symmetry are equivalent in the case of
 QCD.
 \subsubsection{Connection to ``sobar" modes}
 \itt
The physical modes, namely the pions and vector mesons in
superdense matter that I may call ``super-pions" and
``super-vectors," respectively, can be viewed as bound states of
gapped diquarks just as the pions and the vectors in nuclear
matter are viewed as bound states of a quark and an antiquark
coupled to particle-hole excitations of the appropriate quantum
numbers. The corresponding particle-hole configurations are called
``sobars." Given an EQCD Lagrangian, one can use Bethe-Salpeter
equation to compute the bound states of pionic and vector-meson
quantum numbers. This has been done [RWZ], with however the sobar
configurations ignored. Doing consistent calculations that involve
sobars remains an open problem in both low density and high
density. Let me present my conjecture of what might happen.

Below the chiral transition density, the sobar configurations are
unimportant at very low densities but become important as density
increases. For instance, the fact that the vector-meson (e.g.,
$\rho$) mass falls to zero in the chiral limit as the critical
density is approached, a theme that will be developed below, is
crucially dependent on the increasing importance of the sobar
configuration in the vector channel [KRBR].

The situation at super-high density above the chiral transition
point seems to be opposite to the above. At the asymptotic
density, the massive vector mesons decouple and the sobars
disappear due to weak coupling. As density goes down, the coupling
becomes strong and sobars start figuring and toward the chiral
transition point, the sobars will dominate in triggering the
vanishing of the super-vector mass emerging as massless gluons.

Thus hidden/explicit local symmetry appears in a similar way in
two regimes, above and below the chiral transition point, but with
an opposite tendency.
 \subsubsection{Comments: continuity/duality and Cheshire Cat}
 \itt
The massive gluons ``ate up" the scalar Goldstones but the octet
psuedoscalar Goldstones remain as physical excitations. These
pseudoscalars have the same quantum numbers as the pion octet in
the zero-density regime. The gluons inherit the same quantum
numbers as the octet vector mesons $\rho$, $\omega$ and $K^\star$.
The quarks become massive due to the gap $\Delta$ and can be
identified, in quantum numbers, with the octet baryons of the
zero-density regime. One way of seeing this is to consider the
baryons as octet solitons or skyrmions -- called qualitons [HRZ]
-- of the effective Lagrangian (\ref{effCFL}). Thus we have an
uncanny one-to-one correspondence of quarks/baryons and
gluons/vector mesons [SW]. I might identify this as another
manifestation of Cheshire Cat.

For completeness, I should briefly comment on the spectrum of the
pseudo-Goldstones when quark mass terms are included although this
is not directly relevant to our discussion. Since
$\la\bar{\psi}\psi\ra=0$ in this scheme, the term linear in the
quark mass $m_q$ is missing in the Gell-Mann-Oakes-Renner mass
formula for the octet pions. Therefore $m_\pi^2$ goes as $m_q^2$.
It turns out that because of this peculiarity, the
pseudo-Goldstone spectrum is inverted, that is, the kaons are
lighter than the pions.
\section{Lecture III: Color-Flavor Locking and Chiral Restoration}
\subsection{EFT from Color-Flavor-Locked Gauge Symmetry}
\subsubsection{Quark-anti-quark condensates}
\itt We go back to zero density and consider an EFT in close
analogy to the CFL theory for superdense matter. This section
follows Wetterich's idea [WETT].

I start by assuming that the color and flavor lock in the
Nambu-Goldstone phase (in zero density) as in the superdense
matter. In other words, we allow the condensates
 \be
\la\chi_{ij}^{\alpha\beta}\ra\equiv \la
\bar{\psi}_{jL}^\beta\psi_{iR}^\alpha\ra&=&\frac{1}{\sqrt{6}}\chi_0
(\delta_i^\alpha\delta_j^\beta-\frac 13
\delta_{ij}\delta^{\alpha\beta})\nonumber\\
&+&
\frac{1}{\sqrt{3}}\sigma_0\delta_{ij}\delta^{\alpha\beta}.\label{scondensate}
 \ee
The first term on the RHS of (\ref{scondensate}) is a color-octet
condensate and the second term a color-singlet condensate with the
constants $\chi_0$ and $\sigma_0$ representing the magnitude of
the condensates. We are familiar with the singlet condensate
$\sigma_0$ which is usually the only condensate invoked in the
literature. What is not familiar is the color-octet condensate
which has usually been assumed to be zero. Wetterich argues that
instanton interactions in the presence of the octet condensates
can self-consistently generate the attraction. Let me simply
continue assuming that it is non-zero and suggest that there is a
compelling \`a posteriori reason why it should be non-zero.

The octet condensate implies the symmetry breaking pattern
identical to the CFL considered at an asymptotic density in terms
of diquark condensation. Clearly chiral symmetry is broken,
 \be
SU(3)_L\times SU(3)_R\rightarrow SU(3)_V.
 \ee
Furthermore the global color symmetry is also broken as
 \be
SU(3)_C\times SU(3)_V\rightarrow SU(3)_{P}
 \ee
where $P=C+V$ as before. Following the same reasoning given above,
the quarks turn into the baryons we know and love, $p, n, \Sigma,
\Lambda, \Xi$, and the gluons get Higgsed to turn into the vector
mesons $\rho, K^\star, \omega$. Thus the gluon/meson and
quark/baryon continuity pictures repeat here. Now to write the
corresponding EFT, write
 \be
\chi_{ij}^{\alpha\beta}=
\frac{1}{\sqrt{6}}\left\{\xi_{Ri}^\alpha(\xi_{Lj}^\beta)^\star
-\frac 13 U_{ij}\delta^{\alpha\beta}\right\}.
 \ee
Here as in (\ref{condense}), $(\alpha,\beta)$ stand for color
labels and $(i,j)$ for flavors and $U=\xi_L^\dagger\xi_R$. I will
come back to the baryons later and focus here on the mesons. Again
in an arbitrary gauge, we have an octet of pseudoscalars $\pi$, an
octet of scalars $s$ and an octet of vectors $V_\mu$. The
parameters that figure in the corresponding EFT Lagrangian are
again the gauge coupling g, the pion decay constant $F_\pi$ and
$a$ which would be given in terms of the condensates $\chi_0$ and
$\sigma_0$ and those connected to the quark mass terms. The EFT
Lagrangian is of the same form as (\ref{effCFL}),
 \ba
\calL^{eff}_{CFL}=F^2_\pi\Tr(\hat{\alpha}_{A\mu})^2+a F^2_\pi\Tr
(\hat{\alpha}_{V\mu})^2 -\frac{1}{2g^2}\Tr (V_{\mu\nu})^2+\cdots
\label{effzerod}
 \ea
with $V_{\mu\nu}=\del_\mu V_\nu -\del_\nu V_\mu -i[V_\mu, V_\nu]$.
The ellipsis stands for possible higher order terms. The EW fields
can be simply included in generalizing the covariant derivative in
the standard way.

It should be remarked that here the vector $V_\mu$ comes from the
Higgsed gluon ``dressed" by the cloud of nonlinear pions which I
will write down more explicitly below. As such when the vector
gets undressed (by density as will be done below or equivalently
by temperature), it will return to the gluon living in QCD. This
point was made in [BR:PR96] in connection with the possible relay
between an induced flavor gauge symmetry of HLS and fundamental
color gauge symmetry of QCD that was conjectured some years ago.
Again, in the Nambu-Goldstone mode, the vectors are massive by
Higgs mechanism, so the mass is given by
 \be
m_V^2=aF_\pi^2 g^2,
 \ee
i.e., the KSRF-type relation. Formulated this way, although
manifested highly nonlinearly, {\it the local gauge symmetry is
evident} here as in the case of the superdense matter. In fact one
would get to the same answer if one were to start with hidden
local symmetry based on the gauge equivalence between the
nonlinear $G/H$ representation and $G_{global}\times H_{local}$.

So far I have focused on the mesons only. As mentioned, the
baryons also arise from the quarks in the same way as in the
superdense case, namely ``dressed" with pion clouds. Again one can
imagine them arising as skyrmions from the EFT Lagrangian
(\ref{effzerod}).
\subsubsection{Relations between ($F_\pi$, $a$) and ($\chi_0$,
$\sigma_0$)}
 \itt
Wetterich derived several important relations between the
parameters of the EFT and the condensates. Since the discussion is
quite technical, I won't go into detail but quote some of them
here. By writing
 \be
\xi_{L(R)i}^\alpha=[\xi_{L(R)}v]_i^a, \ \ U=\xi_L^\dagger\xi_R
 \ee
in terms of a color-singlet $\xi$ field and a $v\in SU(3)_C$, both
of which are unitary, he relates the vector meson fields $V_\mu$
and the baryon fields $B$, to the gluon fields $A_\mu$ and the
quark fields $\psi$ as
 \be
B&=&Z_\psi^{1/2}\xi^\dagger \psi v^\dagger,\nonumber\\
V_\mu&=&v (A_\mu  +\frac{i}{g} \del_\mu) v^\dagger\label{dressed}
 \ee
where $Z_\psi$ is the quark wave function renormalization
constant. Then under the action of $SU(3)_P$ given by an hermitian
$3\times 3$ matrix $\theta_P (x)$,
 \be
\delta B&=& i[\theta_P, B], \nonumber\\
\delta V_\mu &= &i[\theta_P, V_\mu]+\frac{1}{g}\del_\mu
\theta_P,\nonumber\\
\delta \xi_{L,R}&=& -i\xi_{L,R}\theta_P,\nonumber\\
\delta U&=& 0,\nonumber\\
\delta v &=& i\theta_P v.
 \ee
The relations Wetterich obtained are
 \ba
F_\pi&=&2(\sigma_0^2 +\frac 49 \chi_0^2),\\
a&=& \frac{16}{9} (1+x)/x,\\
m_V&=&\sqrt{a}g F_\pi
 \ea
where $x=4\chi_0^2/(9\sigma_0^2)$. By fixing $\sigma_0$ and
$\chi_0$ from the experimental values of $F_\pi=93$ MeV and
$m_\rho=770$ MeV, he obtained
 \ba
a\approx 1.7
 \ea
which should be compared with the KSRF relation that is given for
$a=2$. In fact, he recovers all the results given by HLS theory
including the vector dominance with $g_{\gamma\pi\pi}\approx 0$.

What seems surprising in the Wetterich's result is that the pion
decay constant is dominated by the octet condensate $\chi_0$, in
contrast to the conventional thinking that it is related entirely
to the singlet condensate $\sigma_0$. Why this is so and whether
this is not in conflict with QCD proper is yet to be clarified.
The vector meson mass is also dominated by the octet condensate
whereas the baryon mass is given by a combination of the octet and
singlet condensates with the latter being more important.
\subsection{EFT from Hidden Local Flavor Symmetry}
\itt The Bando-Kugo-Yamawaki (BKY in short) hidden gauge symmetry
theory has been discussed extensively in review articles, so I
won't spend much space on it. Let me briefly summarize what is in
it.

The reasoning involved here is a ``bottom-up" one whereas the CFL
strategy is a ``top-down" one. One starts with the observation
that a nonlinear theory with the coordinates in the coset space
$G/H$ where $G=SU(3)_L\times SU(3)_R$ is the unbroken symmetry and
$H=SU(3)_{V=L+R}$ is the invariant subgroup with the remaining
unbroken symmetry is ``gauge equivalent" to the linear theory with
$G_{global}\times H_{local}$. If one takes the unitary field $U
\in G/H$ as
 \be
U(x)=\xi_L^\dagger (x)\xi_R (x)
 \ee
with the transformation $U\rightarrow g_L U g_R^\dagger$, there is
a hidden local symmetry
 \be
\xi_L^\dagger (x)\xi_R (x)=\xi_L^\dagger (x)h^\dagger (x)
h(x)\xi_R^\dagger (x),
 \ee
 \be
\xi_L&\rightarrow& h(x)\xi_L g_L^\dagger,\nonumber\\
\xi_R&\rightarrow& h(x)\xi_R g_R^\dagger
 \ee
with $h\in SU(3)_V$. Now gauging this symmetry with a gauge field
$V_\mu\in SU(3)_V$ and adding a kinetic energy term gives the
hidden local symmetry Lagrangian identical to what we have written
down twice already, e.g., (\ref{effCFL}) and (\ref{effzerod}). The
kinetic term can be ``induced" by quantum loops as in the
$CP^{N-1}$ model. One way of looking at the hidden gauge structure
is to view it as an ``emergent symmetry."
\subsection{Explicit vs. Hidden Gauge Symmetry}
\itt We have seen that the same vector mesons can be viewed from
the point of view of QCD as ``dressed" gluons, and hence bosons in
explicit gauge symmetry that is spontaneously broken and from the
point of view of EFT as $induced$ vectors in hidden gauge symmetry
that is also spontaneously broken. They are the same objects, one
QCD-ish and the other with no apparent relation to QCD {\it
proper}.

An extremely interesting question is whether the two can be
connected. I will discuss how the connection can be made by
matching at a given scale but one can ask a more general question
at this stage and that is: can one always generate a gauge theory
out of non-gauge theoretic model as is done in the present case?
The answer turns out to be yes as Weinberg has discussed.

While the above question is interesting in general in connection
with various dualities in field theory, it has a direct
ramification on the behavior of vector mesons in dense/hot medium.
I will discuss this matter in some detail. But let me here
summarize Weinberg's discussion [WEIN:GAUGE].

Following Weinberg, suppose that one has a theory invariant under
a gauge group $G$ with various matter multiplets in various
representations of G and that the gauge symmetry is spontaneously
broken. The gauge bosons will be Higgsed. One may pick the unitary
gauge and integrate out the massive gauge bosons. This will give
rise to a local field theory that is perturbatively renormalizable
in the sense described above. The resulting effective field theory
will have no hint of the original gauge invariance. Furthermore by
allowing arbitrary gauge invariant interactions in the original
theory, one can obtain a {\it completely general theory} of the
remaining fields. Conversely ``{\it out of any effective field
theory with no gauge symmetry and possibly no global symmetry, we
can obtain a theory with any broken gauge symmetry}". This means
that the effective theory obtained from broken gauge symmetry
cannot have a $unique$ predictive power {\it unless the gauge
coupling is weak}. Weinberg illustrates the above point by showing
that a theory which has no apparent symmetries, given by, say, the
action
 \be
S[\psi]=-\int d^4x\sum_i \bar{\psi}_i\gamma^\mu\del_\mu\psi_i
-G[\psi]
 \ee
where $\psi_i$ is the $i$th component of the Dirac fermion and
$G[\psi]$ is an arbitrary local functional of $\psi$ with no
apparent symmetries, is $equivalent$ to the gauge invariant action
  \be
S[\psi,A,\phi]&=&-\frac{M^2}{2}\int d^4x |\del_\mu\phi -i A_\mu
\phi|^2 -\int\sum_i\bar{\psi}_i\gamma^\mu (\del_\mu -iq_i
A_\mu)\psi_i\nonumber\\
&& - \frac 14\int d^4x F_{\mu\nu}F^{\mu\nu} + G^\prime
[\psi^\prime]
  \ee
where $\phi$ is a scalar field constrained to $|\phi|^2=1$ and
$\psi_i^\prime\equiv \psi_i \phi^{-q_i}$. The gauge transformation
is
  \be
\psi_i (x)\rightarrow e^{iq_i\alpha (x) \psi_i (x)}, \
\phi(x)\rightarrow e^{i\alpha (x)}\phi (x), \ A_\mu (x)\rightarrow
A_\mu (x) +\del_\mu\alpha (x).
 \ee

Doing quantum theory with these two actions is another matter. One
version of the theory may be defined in a certain restricted
domain whereas the other may be valid in a different region.
Treated within the regions in which they are well-defined, they
could give completely different results.

That certain gauge symmetries can arise from non-gauge symmetric
theories makes one think of having ``fundamental theories"
emerging from effective theories. In fact, there are attempts to
``derive" -- as in condensed matter physics -- high-energy
fundamental theories as ``emergent" theories as opposed to the
reductionism that views the Standard Model, gravity etc. as coming
down from M theory. (See the articles by Volovik, Laughlin ...)
Some of the symmetries are logically emerging as for instance
certain abelian and non-abelian gauge symmetries arising as Berry
potentials (see the monograph [NRZ])

I stress the above point particularly in the context of recent
efforts to describe how hadrons (in particular vector mesons)
behave in medium with various effective hadronic models with no or
partial global symmetries. Since the computation is done in a
regime where the perturbative approximations are often not
justifiable, there is no guarantee that the predictions made
perturbatively give the right answer. More specifically consider
the properties of the vector mesons ($\rho$, $\omega$, $a_1$ etc.)
in dense and hot matter as one approaches the chiral transition
point, a subject that I will focus on below. In hot and dense
matter, the vector-meson degrees of freedom need to be considered
explicitly, so the chiral Lagrangian consisting of pions only
which has a systematic chiral expansion has to be implemented with
the vector mesons. Therefore people use various forms of
vector-meson-implemented chiral Lagrangians to investigate hadron
properties in medium. For instance, one typically uses gauged
linear sigma model, nonlinear sigma models suitably ``externally
gauged," or HLS theory etc. In very dilute and low temperatures,
one could use these Lagrangians by organizing a systematic power
expansion scheme and may obtain reasonable results. However as one
increases density/temeperature approaching the chiral restoration
point where the QCD degrees of freedom become explicit, the
fluctuations one calculates need not yield unique results unless
the QCD gauge symmetry is imposed. That is because the variety of
effective theories that have flavor gauge fields at low
density/temperature, with or without flavor gauge invariance, can
``flow" as the scale is dialled in a variety of different
directions and some or most of those directions may not have
anything to do with QCD proper. The variety of symmetries present
in the original effective theories, hidden, mended or explicit,
can manifest in different ways as the symmetry restoration regime
is reached. For instance, the $\rho$ and $a_1$ may or may not
appear as degenerate multiplets, vector dominance may or may not
be operative near the critical point, the vector meson masses may
or may not go to zero in the chiral limit etc. This is the reason
why so many different behaviors are predicted by different people
as to how the vector mesons behave near the chiral transition
point. This means that in order to make sensible predictions, the
theory has to be constrained by the color gauge theory of QCD, a
point further elaborated on in the following discussion.

Before getting into a detailed discussion, I should mention here
that the presence of local gauge symmetry allows systematic higher
order calculations that the theories without cannot. The HLS
Lagrangian that we have here will turn out to enable us to
calculate higher order terms that are consistent with QCD in the
sense that it matches with QCD at some given scale. If one has
various vector mesons (e.g., $\rho$, $a_1$ etc.) that have no
gauge invariance, then higher order terms computed in such
theories cannot be controlled and hence cannot be trusted.
\subsection{Vector Manifestation of Chiral Symmetry}
\subsubsection{HLS and chiral perturbation theory} \itt Although I
have written down the broken gauge theory Lagrangian in both ways,
explicit and hidden, I will simply refer to the given Lagrangian
as {\it HLS Lagrangian}.

At the tree level, there are a variety of ways of introducing
vector mesons into the chiral Lagrangian that give the same
results. However beyond the tree level, they are not equivalent.
Some have no meaning at loop order. Others, while loops can be
considered, cannot be controlled. They can give different results
at loop levels even if the loop consideration is justified. The
power of HLS vs. external gauging is that a systematic chiral
expansion is possible in the scheme. In fact, as one nears the
chiral phase transition, this is the only theory that has sensible
higher order terms that are calculable [HY:PR]. To refresh the
memory of the readers, I write down the full leading order
(${\calO}(p^2)$) HLS Lagrangian explicitly:
 \ba
\calL_{HLS}=F^2_\pi\Tr(\hat{\alpha}_{A\mu})^2+ F^2_\sigma\Tr
(\hat{\alpha}_{V\mu})^2 -\frac{1}{2g^2}\Tr (V_{\mu\nu})^2
+\calL_{p^4}. \label{hls}
 \ee
with
 \be
\frac{F^2_\sigma}{F^2_\pi}=a
 \ee
 and
 \be
V_{\mu\nu}&=&\del_\mu V_\nu -\del_\nu V_\mu -i[V_\mu,
V_\nu],\nonumber\\
\hat{\alpha}_{V\mu} (x)&\equiv& (\calD_\mu \xi_L\cdot\xi_L^\dagger
+\calD_\mu \xi_R\cdot\xi_R^\dagger) (-i/2),\nonumber\\
\hat{\alpha}_{A\mu} (x)&\equiv& (\calD_\mu \xi_L\cdot\xi_L^\dagger
-\calD_\mu \xi_R\cdot\xi_R^\dagger) (-i/2)
 \ee and
 \be
\calD_\mu \xi_L&=&\del_\mu-iV_\mu\xi_L+i\xi_L{\cal L}_\mu,\nonumber\\
\calD_\mu \xi_R&=&\del_\mu-iV_\mu\xi_R+i\xi_R{\cal R}_\mu
 \ee
where ${\cal L}_\mu$ and ${\cal R}_\mu$ are respectively left and
right external gauge fields. I put the external gauge fields here
since I will need them in considering correlators later. I  have
not put the mass term in (\ref{hls}) but it should be added for
practical calculations. For the calculations to be described
below, we need the Lagrangian of ${\calO} (p^4)$.  If one includes
the external gauge fields, there are some 35 terms in
$\calL_{p^4}$, the complete list of which can be found in the
Phys. Rep. review of Harada and Yamawaki [HY:PR]. We won't write
it here in full. What we will need for our purpose are the ones
that enter in the vector and axial-vector correlators and will be
written down as needed.

As is done in chiral perturbation theory with the standard chiral
Lagrangian without the vector mesons, we count
 \be
\del_\mu \sim {\cal L}_\mu\sim {\cal R}_\mu \sim {\calO}(p).
 \ee
Here and in what follows, $p$ represents the characteristic small
probe momentum involved. In the same counting the pion mass term
will be $m_\pi^2 \sim \calO (p^2)$ as the leading term in
(\ref{hls}). It is in dealing with the vector mesons that one
encounters an unconventional counting. As pointed out first by
Georgi [GEORGI], in order to have a systematic chiral expansion
with (\ref{hls}), the vector meson mass has to be counted as
 \be
m^2_\rho\sim m_\pi^2\sim {\calO} (p^2).\label{vmass}
 \ee
Since $m^2_\rho\sim g^2 f_\pi^2$, this means that we have to count
$g\sim {\calO} (p)$. Thus we should have
 \be
V_\mu\sim g\rho_\mu\sim {\calO} (p).
 \ee
This is how the vector kinetic energy term in (\ref{hls}) is of
${\calO} (p^2)$.

It might surprise some readers to learn that the vector meson mass
has to be considered as ``light" when in reality it is more than
five times heavier than the pion mass in the vacuum. Even though
one might argue that the expansion is $m_\rho^2/(4\pi F_\pi)^2\sim
0.4$, this is not so small. Surprisingly enough, once one adopts
this counting of the vector mass and the vector field, one can do
a chiral perturbation calculation [HY:MATCH] that is equivalent to
the classic calculation of Gasser and Leutwyler [GASS].  At this
point it is important to note that while one may doubt the
validity of the counting rule in the vacuum, it will however be
fully justified in the scenario where the vector meson mass {\it
does} become comparable to the pion mass as density approaches
that of chiral restoration. So in some sense this counting rule
which is valid at high density is being extrapolated down smoothly
to the zero density regime.

If the Lagrangian (\ref{hls}) is to be taken as an effective field
theory of QCD, then it can make sense only if the parameters of
the Lagrangian are $bare$ parameters defined at a certain given
scale. It is natural to define them at the chiral scale
$\Lambda_\chi$ and that above $\Lambda_\chi$, it is QCD proper
that is operative. Thus we are invited to match the HLS Lagrangian
(\ref{hls}) to QCD at $\Lambda_\chi$ to determine the parameters.
This matching will provide the $bare$ Lagrangian in the Wilsonina
sense with which one can do quantum theory by ``decimating" down
to zero energy/momentum. The matching can be done typically with
correlators in Euclidean space. We do this with the vector and
axial-vector correlators defined as
 \be
i\int d^4x e^{iqx}\la 0|TJ_\mu^a (x)J_\nu^b (0)|\ra &=&\delta^{ab}
\left(q_\mu q_\nu -g_{\mu\nu} q^2\right) \Pi_V (Q^2),\nonumber\\
i\int d^4x e^{iqx}\la 0|TJ_{5\mu}^a (x)J_{5\nu}^b (0)|\ra
&=&\delta^{ab} \left(q_\mu q_\nu -g_{\mu\nu} q^2\right) \Pi_A
(Q^2)
 \ee
with
 \be
Q^2\equiv -q^2.
 \ee
We first define the scale $\bar{\Lambda}$ at which the matching
will be done. One would like to match the HLS theory and QCD at
the point where both are valid. We suppose that the QCD
correlators can be matched to those of HLS at $\bar{\Lambda}$
close to but slightly below $\Lambda_\chi$. At $\bar{\Lambda}$,
the correlators in the HLS sector are supposed to be well
described by the tree contributions to $\calO (p^4)$ when the
Euclidean momentum $Q\sim \bar{\Lambda}$. To compute the tree
graphs, we need the $\calO (p^4)$ Lagrangian that enters into the
correlators,
 \be
\delta
\calL=z_1\Tr\left[\hat{\calV}_{\mu\nu}\hat{\calV}^{\mu\nu}\right]
+z_2\Tr\left[\hat{\calA}_{\mu\nu}\hat{\calA}^{\mu\nu}\right]+
z_3\Tr \left[\hat{\calV}_{\mu\nu} V^{\mu\nu}\right]
 \ee
where $\hat{\calV}_{\mu\nu}$ and $\hat{\calA}_{\mu\nu}$ are
respectively the external vector and axial-vector field tensors
and $V_{\mu\nu}$ is the field tensor for the HLS vector defined
above. The correlators in the HLS sector then are
 \be
\Pi^{(HLS)}_A (Q^2)&=&\frac{F_\pi^2 (\bLambda)}{Q^2}-2 z_2
(\bLambda),\\
\Pi^{(HLS)}_V (Q^2)&=&\frac{F_\sigma^2 (\bLambda)}{M_v^2
(\bLambda)+Q^2}\left[1-2g^2(\bLambda)z_3 (\bLambda)\right]-2z_1
(\bLambda)
 \ee
with
 \be
 M_v^2 (\bLambda)\equiv g^2(\bLambda)F_\sigma^2 (\bLambda).
 \ee
Since we are at the matching scale, there are no loops, i.e., no
flow.

Next we have to write down the correlators in the QCD sector. We
assume that these are given by the OPE to $\calO (1/Q^6)$,
 \be
\Pi^{(QCD)}_A&=&\frac{1}{8\pi^2}\left[-(1+\alpha_s/\pi)\ln\frac{Q^2}{\mu^2}
+\frac{\pi^2}{3}\frac{\la\frac{\alpha_s}{\pi}G_{\mu\nu}
G^{\mu\nu}\ra}{Q^4}
+\frac{\pi^3}{3}\frac{1408}{27}\frac{\alpha_s\la\bar{q}q\ra^2}{Q^6}\right],\\
\Pi^{(QCD)}_V&=&\frac{1}{8\pi^2}\left[-(1+\alpha_s/\pi)\ln\frac{Q^2}{\mu^2}
+\frac{\pi^2}{3}\frac{\la\frac{\alpha_s}{\pi}G_{\mu\nu}
G^{\mu\nu}\ra}{Q^4}
-\frac{\pi^3}{3}\frac{896}{27}\frac{\alpha_s\la\bar{q}q\ra^2}{Q^6}\right]
 \ee
where $\mu$ here is the renormalization scale of QCD (e.g., in the
sense of dimensional regularization)~\footnote{This $\mu$ is not
to be confused with the chemical potential also denoted $\mu$. To
avoid the confusion, I will use $\M$ for the renormalization scale
in what follows.}. Note that the QCD correlators separately depend
explicitly on the renormalization scale $\mu$ but the difference
does not. Such dependence must be lodged in the $z_{1,2}$ terms in
the HLS sector. The condensates and the gauge coupling constant
$\alpha_s$ of course depend implicitly on the scale $\mu$. In
matching the QCD correlators to the HLS' ones, the natural thing
to do is to take $\mu\sim \bLambda$.

The matching is done by equating the correlators and their
derivatives at $Q^2=\bLambda^2$. This gives effectively three
equations. The parameters to be fixed are $F_\pi$, $g$, $a$, $z_3$
and $z_2-z_1$. That makes five parameters. To determine all five,
the on-shell pion decay constant $F_\pi (0)=93$ MeV and the vector
meson mass $m_\rho=770$ MeV are used as inputs. Given $\mu$, one
can then fix all five constants at the scale $\bLambda$. The HLS
Lagrangian with the parameters so determined at $\bLambda$ when
implemented with RGE's with certain quadratic divergences (to be
described below) gives results that are very close to those
obtained by the Gasser-Leutwyler Lagrangian [GASS] that includes
$\calO (p^4)$ counter terms. This means that the counting rule
used for the vector meson is consistent with chiral perturbation
theory. Just to illustrate the point, let me give some numerical
predictions given by the theory. With $\bLambda=1.1\sim 1.2$ GeV,
$\Lambda_{QCD}=400$ MeV, the results are:
 \be
g_\rho&=& 0.116\sim 0.118 \ (0.118\pm 0.003),\nonumber\\
g_{\rho\pi\pi}&=&5.79\sim 5.95 \ (6.04\pm 0.04), \nonumber\\
L_9 (m_\rho)&=&7.55\sim 7.57 \ (6.9\pm 0.7), \nonumber\\
L_{10} (m_\rho)&=&
-7.00\sim -6.23 \ (-5.2\pm 0.3), \nonumber\\
 a(0)&=&1.75\sim 1.85 \
(2)
 \ee
where the numbers in the parenthesis are experimental except for
$a$ which corresponds to VDM (or KSRF) value. For more details,
see the Harada-Yamawaki review [HY:PR].
\subsubsection{The ``vector manifestation (VM)" fixed point}
\itt The HLS theory (\ref{hls}) has a set of fixed points when the
scale is dialled. Among them there is only one fixed point called
``vector manifestation (VM)" that matches with QCD. In the next
subsection, I will show that that fixed point is relevant to
chiral restoration when nuclear matter is compressed or heated.

We start with (\ref{hls}) defined at the scale $\bLambda$, below
which the relevant degrees of freedom are the pions $\pi$ and the
vector mesons $\rho$ (and $\omega$). We would like to do the EFT
as defined above by decimating downward from the scale $\bLambda$.
The parameters $F_\pi$, $g$, $F_\sigma$ (or $a$) and $z_i$ will
flow as the renormalization scale $\M$ is dialled~\footnote{I am
using the symbol $\calM$ as the renormalization scale, reserving
the conventional notion $\mu$ for chemical potential.}. To do
this, one has to choose the gauge and the convenient gauge is the
background gauge. These have been worked out in the background
gauge [HY:PR] and to one-loop order, take the forms~\footnote{It
is not transparent in this form but there is an important
quadratic divergence in the pion loop contributions, hence in
$F_\pi$ (i.e., $X$) as well as in $a$ that plays a crucial role
for chiral restoration. The fixed point in $X$ is governed by this
divergence.}
 \be
\M \frac{dX}{d\M} &=& (2-3a^2G)X -2(2-a)X^2,
\nonumber\\
\M \frac{da}{d\M} &=&-
(a-1)[3a(1+a)G-(3a-1)X],\nonumber\\
\M\frac{d G}{d \M}&=&
-\frac{87-a^2}{6}G^2,\nonumber\\
\M\frac{dz_1}{d\M}&=& C\frac{5-4a+a^2}{12},\nonumber\\
\M\frac{dz_2}{d\M}&=& C\frac{a}{6},\nonumber\\
\M\frac{dz_3}{d\M}&=& C\frac{1+2a-a^2}{6} \label{rgehls}
 \ee
where
 \be
X(\M)\equiv C\M^2/F_\pi^2 (\M), \ \ G(\M)=C g^2, \ \ C =
N_f/\left[2(4\pi)^2\right].
 \ee
I should note that these are Wilsonian renormalization group
equations as defined in my second lecture valid above the vector
meson mass scale $m_\rho$ defined by the on-shell condition
$m_\rho^2=a(m_\rho) g^2 (m_\rho) F_\pi^2 (m_\rho)$ at
$\M=m_\rho$. This means that Equations (\ref{rgehls}) are valid
for $a(\M)G(\M) \leq X(\M)$ but modified below by the decoupling
of the vector mesons.

The fixed points are given by setting the RHS of (\ref{rgehls})
equal to zero. It is found in [HY:FATE] that there are three fixed
points in the relevant region with $a>0$ and $X>0$:
 \be
(X^\star, a^\star, G^\star)= (1,1,0), (\frac 35,\frac 13,0),
\left(\frac{2(2+45\sqrt{87})}{4097},\sqrt{87},\frac{2(11919-176\sqrt{87})}
{1069317}\right)
 \ee
and a fixed line
 \be
(X^\star, a^\star, G^\star)=(0,{\rm any},0).
 \ee
Depending upon how the parameters are dialled, there can be a
variety of flows as the scales are varied. The flows are
interesting for the model {\it per se}. However not all of them
are relevant to the physics given by QCD. In fact, if one insists
that the vector and axial-vector correlators of HLS theory are
matched at the chiral phase transition that is characterized by
the on-shell pion decay constant $F_\pi (\M=0)=0$ to those of QCD
at a scale $\bLambda\sim \Lambda_\chi$, then {\it there is only
one fixed point to which the theory flows and that is the vector
manifestation (VM) fixed point}
 \be
(X^\star, a^\star, G^\star)_{VM}= (1,1,0).
 \ee
This means that independently of what triggers the phase
transition, at the chiral restoration point, the theory must be at
the point $a=1, g=0, F_\pi=0$. As emphasized by Harada and
Yamawaki, this is not Georgi's ``vector limit" [GEORGI] where
$a=1, g=0$ but $F_\pi\neq 0$.

We will see next that density or temperature does indeed drive the
system to the VM fixed point and as a consequence, the vector
meson mass must vanish at the chiral restoration point at least in
the chiral limit. A corollary to this is that if an effective
theory that has all the low-energy symmetries consistent with QCD
but is pnot matched to QCD at an appropriate matching scale can
flow in density/temperature in a direction that has nothing to do
with QCD. In such a model, the vector meson mass need not vanish
at the chiral transition point.
\subsection{The Fate of the Vector Meson in Hot/Dense Matter}
 \itt
In this part of the lecture, I will show that indeed, density or
temperature can drive nuclear matter toward the VM fixed point and
as a consequence, the vector meson mass must vanish at the
critical point at least in the chiral limit. See [HKR]. Since the
temperature case is quite similar to the density case, I will
consider here only the density case. It is of course more relevant
for studying the structure of compact-star matter. The same
conclusion holds for high temperature case relevant to heavy-ion
physics.
\subsubsection{Renormalization group equations for dense matter}
\itt Consider a many-body hadronic system at a density $n$ or
equivalently chemical potential $\mu$ ($n$ and $\mu$ will be used
interchangeably). In the presence of matter density, the system
loses Lorentz invariance and hence the theory should be formulated
with an $O(3)$ invariance. This means that one has to separate the
time and space components of vector objects (like currents etc.).
It turns out that one can proceed as if we have Lorentz invariance
and at the end of the day, make the distinction when necessary.
The details of non-Lorentz invariant structure are given in the
paper [HKR].

The Lagrangian density (\ref{hls}) including the $\calO (p^4)$
term contains no fermions and hence is not sufficient.  To
introduce fermion degrees of freedom, we assume that as one
approaches the chiral transition point from below, the quasiquark
(or constituent quark) description is more appropriate than
baryonic. Denoting the quasiquark by $\psi$, we write the fermion
part of the Lagrangian as
\begin{eqnarray}
 \delta {\cal L}_{F} &=& \bar \psi(x)( iD_\mu \gamma^\mu
       - \mu \gamma^0 -m_q )\psi(x)\nonumber\\
 &&+ \bar \psi(x) \left(
  \kappa\gamma^\mu \hat{\alpha}_{\parallel \mu}(x )
+ \lambda\gamma_5\gamma^\mu \hat{\alpha}_{\perp\mu}(x) \right)
        \psi(x) \label{lagbaryon}
 \end{eqnarray}
where $D_\mu\psi=(\partial_\mu -ig\rho_\mu)\psi$ and $\kappa$ and
$\lambda$ are constants to be specified later. The HLS Lagrangian
we work with is then given by (\ref{hls}) and (\ref{lagbaryon}).

We consider matching at the scale $\bLambda$. In the presence of
matter, the matching scale will depend upon density. We use the
same notation as in the vacuum with the $\mu$ dependence
understood. At the scale $\bLambda$, the correlators are given by
the tree contributions and hence by (\ref{hls}) only. Since there
is no flow, (\ref{lagbaryon}) which can figure only at loop order
does not enter. We have
\begin{eqnarray}
\Pi_A^{\rm(HLS)}(Q^2) &=& \frac{F_\pi^2(\bLambda;\mu)}{Q^2} - 2
z_2(\bLambda;\mu) \ ,
\nonumber\\
\Pi_V^{\rm(HLS)}(Q^2) &=& \frac{
  F_\sigma^2(\bLambda;\mu)
  \left[ 1 - 2 g^2(\bLambda;\mu) z_3(\bLambda;\mu) \right]
}{
  M_\rho^2(\bLambda;\mu) + Q^2
}  - 2 z_1(\bLambda;\mu) \ , \label{corrHLS}
\end{eqnarray}
where $M_\rho^2(\bLambda;\mu) \equiv g^2(\bLambda;\mu)
F_\sigma^2(\bLambda;\mu)$ is the bare $\rho$ mass, and
$z_{1,2,3}(\bLambda;\mu)$ are the bare coefficient parameters of
the relevant ${\cal O}(p^4)$ terms, all at $\M=\bLambda$. As was
done above in the free space, one matches these correlators to
those of QCD and obtain the $bare$ parameters of the HLS
Lagrangian defined at scale $\bLambda$. Since the condensates of
the QCD correlators are density dependent, the $bare$ parameters
of the Lagrangian must clearly be density dependent. This density
dependence in the HLS sector that we will refer to as ``intrinsic
density dependence" can be understood in the following way. First
of all the matching scale $\bLambda$ will have an intrinsic
density dependence. Secondly the degrees of freedom $\Phi_H$
lodged above the scale $\bLambda$ which in full theory are in
interactions with the nucleons in the Fermi sea will, when
integrated out for EFT, leave their imprint of interactions --
which is evidently density-dependent -- in the coefficients of the
Lagrangian. This ``{\it intrinsic density dependence}" is
generally absent in loop-order calculations that employ effective
Lagrangians whose parameters are fixed by comparing with
experiments in the matter-free space. Such theories miss the VM
fixed point.

In the QCD correlators, going to the Wigner phase with $\la
\bar{q}q\ra_{\mu_c}=0$ implies that at $\M=\bLambda$
 \be
\Pi^{(QCD)}_V (Q^2;\mu_c)=\Pi^{(QCD)}_A (Q^2;\mu_c)
 \ee
which implies by matching that
 \be
\Pi^{(HLS)}_V (Q^2;\mu_c)=\Pi^{(HLS)}_A (Q^2;\mu_c).
 \ee
This means from (\ref{corrHLS}) that
\begin{eqnarray}
&& g(\bLambda;\mu) \mathop{\longrightarrow}_{\mu \rightarrow
\mu_c} 0 \ , \qquad a(\bLambda;\mu) \mathop{\longrightarrow}_{\mu
\rightarrow \mu_c} 1 \ ,
\nonumber\\
&& z_1(\bLambda;\mu) - z_2(\bLambda;\mu)
\mathop{\longrightarrow}_{\mu \rightarrow \mu_c} 0 \ .
\label{VMmu}
\end{eqnarray}
Note that this gives no condition for $F_\pi (\bLambda;\mu_c)$. In
fact it is not zero.

Given (\ref{VMmu}) at $\mu=\mu_c$, how the parameters vary as they
flow to on-shell is governed by the RGEs. With the contribution
from the fermion loops given by (\ref{lagbaryon}) added, the RGEs
now read
 \be
\M \frac{dF_\pi^2}{d\M} &=& C[3a^2g^2F_\pi^2
+2(2-a)\M^2] -\frac{m_q^2}{2\pi^2}\lambda^2 N_c\nonumber\\
\M \frac{da}{d\M} &=&-C
(a-1)[3a(1+a)g^2-(3a-1)\frac{\M^2}{F_\pi^2}]
+a\frac{\lambda^2}{2\pi^2}\frac{m_q^2}{F_\pi^2}N_c\nonumber\\
\M\frac{d g^2}{d \M}&=& -C\frac{87-a^2}{6}g^4
+\frac{N_c}{6\pi^2}g^4 (1-\kappa)^2\nonumber\\
\M\frac{dm_q}{d\M}&=& -\frac{m_q}{8\pi^2}[ (C_\pi-C_\sigma)\M^2
-m_q^2 (C_\pi-C_\sigma) +M_\rho^2C_\sigma -4C_\rho
]\label{rgewbaryons}
 \ee
where $C = N_f/\left[2(4\pi)^2\right]$ and
 \be
C_\pi&\equiv &(\frac{\lambda}{F_\pi})^2 \frac{N_f^2 - 1}{2N_f}
\nonumber\\
C_\sigma&\equiv &(\frac{\kappa}{F_\sigma})^2 \frac{N_f^2 -
1}{2N_f}
\nonumber\\
C_\rho&\equiv & g^2 (1-\kappa)^2 \frac{N_f^2 - 1}{2N_f}. \nonumber
 \ee
Quadratic divergences are present also in the fermion loop
contributions as in the pion loops contributing to $F_\pi$. Note
that since $m_q=0$ is a fixed point, the fixed-point structure of
the other parameters is not modified. Specifically, when $m_q=0$,
$(g,a)=(0,1)$ is a fixed point. Furthermore $X=1$ remains a fixed
point. Therefore $(X^\star,a^\star,G^\star,m^\star_q)=(1,1,0,0)$
is the VM fixed point.
\subsubsection{Hadrons near $\mu=\mu_c$}
\itt Let us see what the above result means for hadrons near
$\mu_c$. To do this we define the ``on-shell" quantities
\begin{eqnarray}
&&
  F_\pi = F_\pi(\M=0;\mu) \ ,
\nonumber\\
&&
  g = g\mbox{\boldmath$\bigl($}
  \M = M_\rho(\mu);\mu
  \mbox{\boldmath$\bigr)$}
\ , \quad
  a = a\mbox{\boldmath$\bigl($}
  \M = M_\rho(\mu);\mu
  \mbox{\boldmath$\bigr)$}
\ , \label{on-shell para mu}
\end{eqnarray}
where $M_\rho$ is determined from the ``on-shell condition":
\begin{eqnarray}
 M_\rho^2 = M_\rho^2(\mu) =
  a\mbox{\boldmath$\bigl($}
  \M = M_\rho(\mu);\mu
  \mbox{\boldmath$\bigr)$}
  g^2\mbox{\boldmath$\bigl($}
  \M = M_\rho(\mu);\mu
  \mbox{\boldmath$\bigr)$}
  F_\pi^2\mbox{\boldmath$\bigl($}
  \M = M_\rho(\mu);\mu
  \mbox{\boldmath$\bigr)$}
\ .
\end{eqnarray}
Then, the parameter $M_\rho$ in this paper is renormalized in such
a way that it becomes the pole mass at $\mu=0$.

We first look at the ``on-shell'' pion decay constant $f_\pi$. At
$\mu=\mu_c$, it is given by
 \ba
f_\pi (\mu_c)\equiv f_\pi (\M=0;\mu_c)=F_\pi (0;\mu_c) + \Delta
(\mu_c)
 \ea
where $\Delta$ is dense hadronic contribution arising from fermion
loops involving (\ref{lagbaryon}). It has been shown (see [HKR])
that up to ${\cal O}(p^6)$ in the power counting,
$\Delta(\mu_c)=0$ at the fixed point $(g,a,m_q)=(0,1,0)$. Thus
 \be
f_\pi (\mu_c)=F_\pi (0;\mu_c)=0.
 \ee
This is the signal for chiral symmetry restoration.  Since
 \be
F_\pi^2 (0;\mu_c)=F_\pi^2
(\bLambda;\mu_c)-\frac{N_f}{2(4\pi)^2}\bLambda^2,\label{fpieq}
 \ee
and at the matching scale $\Lambda$, $F_\pi^2 (\Lambda;\mu_c)$ is
given by a QCD correlator at $\mu=\mu_c$ -- presumably measured on
lattice, $\mu_c$ can in principle be computed from
 \ba
  F_\pi^2 (\bLambda;\mu_c)=\frac{N_f}{2(4\pi)^2}\bLambda^2 \ .
 \ea
In order for this equation to have a solution at the critical
density, it is necessary that $F_\pi^2 (\bLambda;\mu_c)/F_\pi^2
(\bLambda;0) \sim 3/5$. We do not have at present a reliable
estimate of the density dependence of the QCD correlator to verify
this condition but the decrease of $F_\pi$ of this order in medium
looks quite reasonable.~\footnote{Since $F_\pi^2 (\bLambda;\mu_c)$
is a slowly varying function of $\bLambda$ and hence too much
fine-tuning will be required, (\ref{fpieq}) does not appear to be
a useful formula for determining $\mu_c$.}

Next we compute the $\rho$ pole mass near $\mu_c$. The calculation
is straightforward, so we just quote the result. With the
inclusion of the fermionic dense loop terms, the pole mass, for
$M_\rho, m_q \ll k_F$ (where $k_F$ is the Fermi momentum), is  of
the form
 \be
 m_\rho^2(\mu) &=& M_\rho^2 (\mu) + g^2
  \, G(\mu)\ ,\label{mrho at T 2}\\
 G(\mu)&=&\frac{\mu^2}{2\pi^2} [\frac 13 (1-\kappa)^2 +N_c
 (N_fc_{V1}+c_{V2})]\ .
  \ee
At $\mu=\mu_c$, we have $g=0$ and $a=1$ so that $M_\rho (\mu)=0$
and since $G(\mu_c)$ is non-singular, $m_\rho=0$. Thus the fate of
the $\rho$ meson at the critical density is as follows: {\it As
$\mu_c$ is approached, the $\rho$ becomes sharper and lighter with
the mass vanishing at the critical point in the chiral limit.} The
vector meson meets the same fate at the critical temperature
[HS:T].

So far we have focused on the critical density at which the
Wilsonian matching clearly determines $g=0$ and $a=1$ without
knowing much about the details of the current correlators. Here we
consider how the parameters flow as function of chemical potential
$\mu$. In low density region, we expect that the ``intrinsic''
density dependence of the bare parameters is small. If we ignore
the intrinsic density effect, we may then resort to
Morley-Kislinger (MS) theorem 2 [MK] which states that given an
RGE in terms of $\M$, one can simply trade in $\mu$ for $\M$ for
dimensionless quantities and for dimensionful quantities with
suitable calculable additional terms. The results are
 \be
\mu \frac{dF_\pi^2}{d\mu} &=& -2F_\pi^2 + C[3a^2g^2F_\pi^2
+2(2-a)\M^2] -\frac{m^2}{2\pi^2}\lambda^2 N_c\nonumber\\
\mu \frac{da}{d\mu} &=&-C
(a-1)[3a(1+a)g^2-(3a-1)\frac{\mu^2}{F_\pi^2}]
+a\frac{\lambda^2}{2\pi^2}\frac{m^2}{F_\pi^2}N_c\nonumber\\
\mu\frac{d g^2}{d \mu}&=& -C\frac{87-a^2}{6}g^4
+\frac{N_c}{6\pi^2}g^4 (1-\kappa)^2\nonumber\\
\mu\frac{dm_q}{d\mu}&=& -m_q -\frac{m_q}{8\pi^2}[
(C_\pi-C_\sigma)\mu^2 -m_q^2 (C_\pi-C_\sigma) +M_\rho^2C_\sigma
-4C_\rho ]
 \ ,
\label{RGE-mu}
\end{eqnarray}
where $F_\pi$, $a$, $g$, etc.~are understood as $F_\pi({\cal
M}=\mu;\mu)$, $a({\cal M}=\mu;\mu)$, $g({\cal M}=\mu;\mu)$, and so
on.

It should be stressed that the MK theorem presumably applies in
the given form to ``fundamental theories" such as QED but not
without modifications to effective theories such as the one we are
considering. The principal reason is that there is a change of
relevant degrees of freedom from above $\bLambda$ where QCD
variables are relevant to below $\bLambda$ where hadronic
variables figure. Consequently we do not expect Eq.~(\ref{RGE-mu})
to apply in the vicinity of $\mu_c$. Specifically, near the
critical point, the {\it intrinsic density dependence} of the bare
theory will become indispensable and the naive application of
Eq.~(\ref{RGE-mu}) should break down. One can see this clearly in
the following example: The condition $g(\M=\mu_c;\mu_c)=0$ that
follows from the QCD-HLS matching condition, would imply, when
(\ref{RGE-mu}) is naively applied, that ${g}(\mu)=0$ for {\it all}
$\mu$. This is obviously incorrect. Therefore near the critical
density the {\it intrinsic density dependence} should be included
in the RGE: Noting that Eq.~(\ref{RGE-mu}) is for, e.g., $g({\cal
M}=\mu;\mu)$, we can write down the RGE for $g$ corrected by the
{\it intrinsic density dependence} as
\begin{eqnarray}
\mu \frac{d}{d\mu} g(\mu;\mu)  \quad = \left.
  {\cal M} \frac{\partial}{\partial {\cal M}} g({\cal M} ;\mu)
\right\vert_{{\cal M} = \mu} + \left.
  \mu \frac{\partial}{\partial \mu} g({\cal M} ;\mu)
\right\vert_{{\cal M} = \mu} \ , \label{RGE-g2}
\end{eqnarray}
where the first term in the right-hand-side reproduces
Eq.~(\ref{RGE-mu}) and the second term appears due to the {\it
intrinsic density dependence}. Note that $g=0$ is a fixed point
when the second term is neglected (this follows from
(\ref{RGE-mu})), and the presence of the second term makes $g=0$
be no longer the fixed point of Eq.~(\ref{RGE-g2}). The condition
$g(\mu_c;\mu_c)=0$ follows from the fixed point of the RGE in
$\M$, but it is not a fixed point of the RGE in $\mu$. The second
term can be determined from QCD through the Wilsonian matching.
However, we do not presently have reliable estimate of the $\mu$
dependence of the QCD correlators. Analyzing the $\mu$ dependence
away from the critical density in detail has not yet been worked
out.

The Wilsonian matching of the correlators at
$\Lambda=\Lambda_\chi$ allows one to see how the $\rho$ mass
scales very near the critical density (or temperature). For this
purpose, it suffices to look at the intrinsic density dependence
of $M_\rho$. We find that close to $\mu_c$
 \be
M_\rho^2 (\Lambda;\mu)\sim \frac{\la\bar{q}q (\mu)\ra^2}{F_\pi^2
(\Lambda;\mu) \Lambda^2}
 \ee
which implies that
 \be
\frac{m_\rho^\star}{m_\rho}\sim
\frac{\la\bar{q}{q}\ra^\star}{\la\bar{q}{q}\ra}.\label{BRscaling}
 \ee
Here the star denotes density dependence.  Note that Equation
(\ref{BRscaling}) is consistent with the ``Nambu scaling" or more
generally with sigma-model scaling. How this scaling fares with
nature is discussed in [BR:PR01].
\subsection{A Comment on BR Scaling}
\itt It appears that the vector meson mass (and that of other
mesons other than pions) scales as
 \be
\frac{m_\rho^\star}{m_\rho}\sim
\left(\frac{\la\bar{q}{q}\ra^\star}{\la\bar{q}{q}\ra}\right)^{1/2}
\label{scalingL}
 \ee
at low density near nuclear matter and as (\ref{BRscaling}) near
the critical point. If one considers baryons as bound states of
quasiquarks near the critical point, then it seems quite
reasonable to conclude that the baryon mass scales like the meson
mass near the chiral transition as found in the BR
scaling.~\footnote{At low density below nuclear mater density, the
nucleon mass $m_N^\star$ (identified below as ``Landau mass"")
scales faster than the vector meson mass because of the extra
factor $\sqrt{\frac{g_A^\star}{g_A}}$ (which is less than one and
reaches a constant at about nuclear matter density).} However as
suggested by Oka et al. [JOKA], there can be a mirror symmetry in
the baryon sector -- which is a sort of ``mended symmetry" in the
sense of Weinberg [WEIN:MEND] -- which makes parity doublets come
together at the chiral restoration point to a non-vanishing common
mass $m_0\sim 500$ MeV~\footnote{I am grateful for discussions
with Makoto Oka on this possiblity.}. In this case, the BR scaling
will not be effective in the baryon sector. At the moment, this
mirror symmetry scenario, somewhat unorthodox it might appear to
be, cannot be ruled out.
\section{Lecture IV: Fermi-Liquid Theory as EFT for
Nuclear Matter}
 \itt
In the previous two lectures, I discussed the most dilute nuclear
system, i.e., two nucleon system, the densest system at infinite
density and then a moderate density system near chiral
restoration. In all cases, fixed points played an important role.
Here I will discuss the matter at normal nuclear matter density,
namely nuclear matter with density $n_0\sim 0.16/{\rm fm}^3$. This
system turns out to be governed by another fixed point called
``Fermi-liquid fixed point" and I will develop the idea that this
fixed point theory can be mapped to an effective chiral field
theory with the parameters of the Lagrangian ``running" with an
{\it intrinsic density dependence} introduced above and that can
correctly describe nuclear matter. The density dependence will
turn out to be that of Brown-Rho (BR) scaling [BR:91] discovered
in a completely different context.
\subsection{Setting Up EFT}
\itt In this lecture, I shall focus on baryonic sector, in
particular the structure of many-baryon systems. In the preceeding
lecture, the issue we addressed was the property of mesons when
the mesons are squeezed by baryonic matter. Here we are addressing
the baryonic system itself, that is, its structure and property.

The first question one could ask is: Given an effective Lagrangian
whose $bare$ parameters are determined by matching with QCD at a
scale $\bLambda\sim \Lambda_\chi\sim 4\pi f_\pi$, how can one
describe nuclear matter with this Lagrangian? This question was
first raised in the context of effective chiral field theory by
Lynn [LYNN] and his initial answer was that nuclear matter arises
as a ``chiral liquid" nontopological soliton. This is somewhat
similar to the idea that nuclear matter is a topological soliton
of the baryon charge $B=\infty$, namely, a skyrmion matter I
mentioned in my first lecture. One might side-step the fundamental
issue of how a chiral liquid arises from effective Lagrangian of
QCD by assuming {\it ab initio} the existence of a Fermi sea and
then calculate fluctuations on top of the Fermi surface by chiral
perturbation theory [KFW].

In this lecture, I will not go into the above fundamental issue
and start with an EFT defined at what I could call ``chiral liquid
scale" $\Lambda_F <\bLambda$ set by the Fermi momentum $k_F$. In
doing this, one can choose $\Phi_L$ in various different ways. For
instance, one can include in $\Phi_L$ : (1) $N, \pi, \rho, \omega
...$; (2) $N, \pi$, no vector mesons; (3) $\pi, \rho, \omega$, no
nucleons; (4) $\pi$, no vector mesons, no nucleons. The
possibilities (3) and (4) rely on topological solitons and (1) and
(2) on non-topological solitons. I will choose (1) and (2) and use
them interchangeably.

\begin{figure}[hbt]
\vskip 0.cm
 \centerline{\epsfig{file=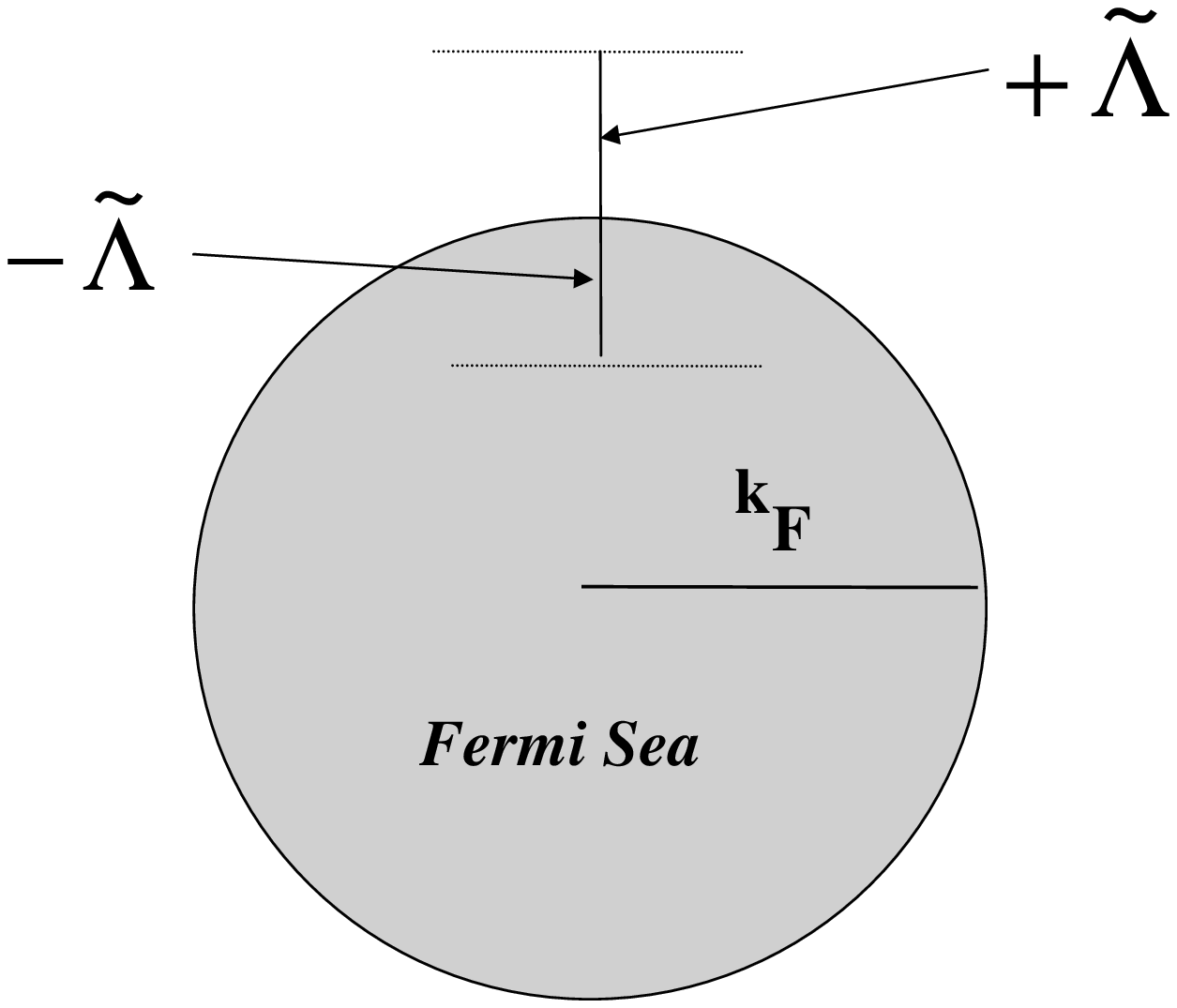,width=9cm,angle=-90}}
\vskip -2.5cm \caption{\small The decimation from
$\tilde{\Lambda}$ on top of the Fermi sea down to the Fermi
surface.}\label{FS}
 \end{figure}

First consider the option (2) which can be thought of as arising
when the vector mesons are all integrated out. And furthermore,
imagine the decimation is effectuated from $\bLambda$ to
$\tilde{\Lambda} <\Lambda_F$. The space we are concerned with can
be visualized as Fig. \ref{FS}.
The $bare$ action that we will start with is supposed to be
determined at $\tilde{\Lambda}$ and takes the form
 \be
S&=& S_0 +S_\pi +S_I,\\
S_0&=& \int dt d^3p \left[N^\dagger (p) i\del_t N(p)-(\epsilon
(p)-\epsilon_F)N^\dagger (p) N(p)\right],\nonumber\\
S_I&=& \int dt \int d^3p_1\cdots d^3p_4 \nonumber\\
&&\delta^3 (\vec{p_1}+\vec{p_2}+\vec{p_3}+\vec{p_4})
u(p_1,p_2,p_3,p_4) N^\dagger (p_1)N(p_2)N^\dagger
(p_3)N(p_4)+\cdots
 \ee
where $S_\pi$ is the well-known pionic action given entirely by
chiral symmetry which I will not write down explicitly here. The
four-Fermi vertex $u$ contains both local and non-local
interactions, the former coming from integrating out the vector
mesons and other heavy mesons and the latter coming from the pion
exchange. When we localize the interactions, then the local part
of the pion exchange will combine to the heavy meson part.

\begin{figure}[hbt]
\vskip 0.cm
 \centerline{\epsfig{file=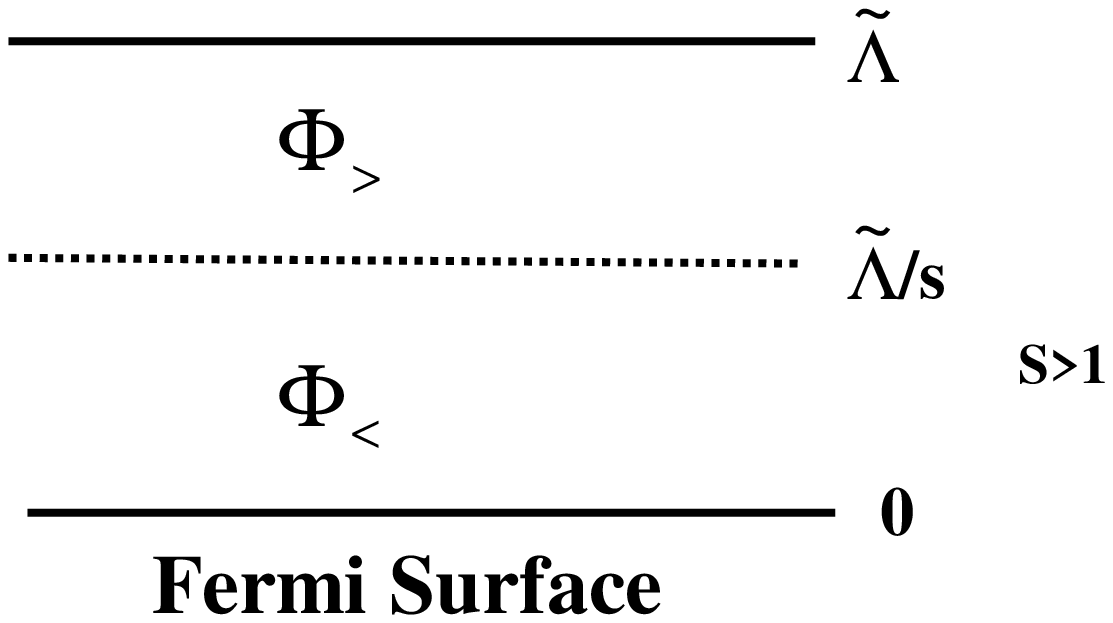,width=10cm,angle=-90}}
\vskip -4.5cm \caption{\small The decimation by integrating out
$\Phi_>$  from $\tilde{\Lambda}$ to $\tilde{\Lambda}/s$ for $s>1$
on top of the Fermi.}\label{decimation}
 \end{figure}

The objective of the calculation is to obtain the $S^{eff}$ of
 \be
Z&=&\int [d\Phi_<] e^{-S^{eff} (\Phi_<)},\\
S^{eff} (\Phi_<)&=&
\int^{\tilde{\Lambda}/s}_{-\tilde{\Lambda}/s}\calL^{eff} (\Phi_<).
 \ee
with $s>1$. The degrees of freedom integrated out, i.e., $\Phi_>$
is indicated in Fig. \ref{decimation}. The effective action
$S^{eff}$ is to be computed in the manner described in the first
part of my second lecture. I will describe what comes out in the
next subsection.
The full decimation corresponds to sending $s$ to $\infty$.
\subsection{Fermi-Liquid Fixed Points}
\begin{figure}[htb]
\vskip -0.0cm
 \centerline{\epsfig{file=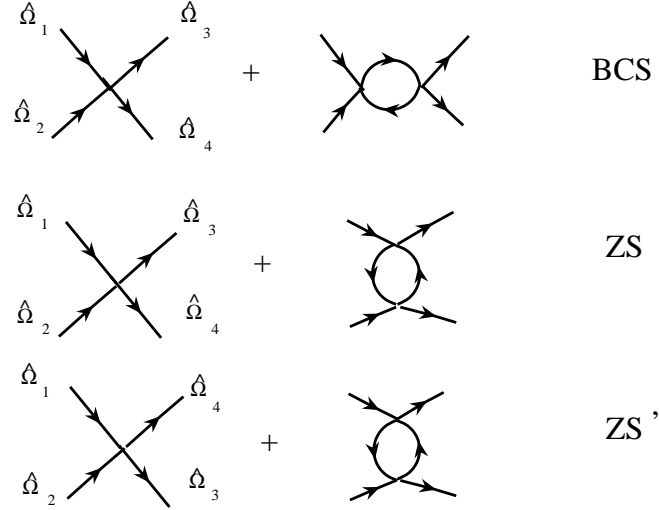,width=12cm,angle=0}}
\vskip -.5cm \caption{\small Tree and one-loop graphs for the BCS
channel $\hO_1+\hO_2=\hO_3+\hO_4=0$, the ZS channel $\hO_1=\hO_3,
\hO_2=\hO_4$ and ZS$^\prime$ channel $\hO_2=\hO_3, \hO_1=\hO_4$
where $\hO$ is the unit vector in 2 dimensions $\hat{\Omega}
=\vec{p}/|\vec{p}|$. Note that $|\vec{p}_i|=k_F$.}\label{graphs}
\end{figure}

\itt By now we know what the standard procedure is. First we
decide that $S_0$ remain marginal under the scaling. This means
that the Fermi velocity $k_F/m^\star$ is marginal where $m^\star$
is the effective mass to be defined precisely below. We choose to
fix the Fermi momentum $k_F$. This choice implies then that {\it
$m^\star$ should be a fixed-point quantity}. This $m^\star$ is the
Landau mass for the quasiparticle in Fermi-liquid theory. Next
once we have the scaling law for the nucleon field by the rule
that $S_0$ be marginal and $m^\star$ be at a fixed point, then we
find that for the component of $u$ that does not scale, i.e., goes
as a constant, the four-Fermi interaction $S_I$ is marginal for
(1) the BCS channel denoted by $V$, (2) the ZS (zero sound)
denoted by $\calF$ and (3) the ZS$^\prime$ denoted by
$\bar{\calF}$ as pictured in Fig. \ref{graphs}. The quaisparticles
interacting through the four-Fermi vertex are sitting on the Fermi
surface, i.e., $|\vec{p_i}|=k_F$ with only the angular variables
varying. (For simplicity, I am using the kinematics in 2 space
dimensions in Fig. \ref{graphs}. The kinematics are similar in 3
space dimensions, just more complicated.)  A constant $u$ is a
local vertex and it will contain then local interactions coming
from heavy-meson exchanges as well as the contact interaction in
the pion exchange. Higher-order terms in momentum transfer that
arise in localizing the non-local terms will be ``irrelevant" and
hence do not contribute in the decimation.

Now given the marginal terms, one needs to compute the loop graphs
in the decimation procedure. How this is done is explained in
detail in the reviews and lecture notes, e.g., [SHANK, POL, CFROS].
Let me summarize:
\begin{itemize}
\item {\bf The BCS channel}:

The tree term is marginal as mentioned above. One-loop as well as
higher-order graphs are non-vanishing and contribute to the
renormaliztion group equation for the interaction. Summing the
graphs to all orders, one finds
 \be
V\sim \frac{v}{1+c v \ln s}
 \ee
where $v$ is the constant interaction coming from the $u$ vertex
in the BCS channel and $c$ is a positive constant. If $v$ is
repulsive, $v>0$, then nothing interesting happens and as $s$ goes
to $\infty$ in decimation, the interaction dwindles away. However
if the interaction is attractive, $v<0$, then there is a Landau
pole and the system becomes unstable. This is the BCS phenomenon,
leading to superconductivity if charged and to superfluidity if
uncharged. This is the phenomenon that takes place at the
superhigh density considered above when the color-flavor locking
takes place, giving rise to color superconductivity.
\item {\bf The ZS channel}:

It is easy to show that the loop terms vanish identically to all
orders. This means a fixed point
 \be
\frac{d\calF}{d\ln s}=0.
 \ee

\item {\bf The ZS$^\prime$ channel}:

Here the loop correction to the RGE goes like $\sim
\frac{d\tilde{\Lambda}}{k_F}\rightarrow 0$ for a given $k_F$ where
$k_F$ is the Fermi momentum. This can be shown to be the case to
all orders of loop corrections. The argument here is like $1/N$
expansion with $\frac{k_F}{d\tilde{\Lambda}}$ playing the role of
$N$. Thus
 \be
\frac{d\bar{\calF}}{d\ln s}=0.
 \ee
\end{itemize}

To summarize, nuclear matter can be described by an EFT with a
Fermi surface with two fixed point quantities, the effective mass
of the quasiparticle $m^\star$ called ``Landau effective mass" and
the interaction $\calF$ called ``Landau quasiparticle
interaction". What is essential in this EFT description is that
one holds fixed the Fermi momentum $k_F$ which is equivalent to
fixing density. This means that {\it for a given density, one has
the sets of Landau parameters that remain invariant under the
renormalization group flow}.
\subsection{``Intrinsic Density Dependence" and Landau Parameters}
\itt In Lecture III, we learned that the HLS theory matched to QCD
at the matching scale $\bLambda$ in dense medium has a set of
parameters that have intrinsic density dependence that is not
accessible to perturbative calculation. Such a theory, when probed
at a given density $n\sim n_0$ ($n_0$ is nuclear matter density),
will then possess density dependent masses and coupling constants.
We have just learned above that the Landau fixed point parameters
are also given for given densities, that is,  density-dependent.
The question I want to pose here is: Is there any connection
between the two?

I do not have the clear answer to this intriguing question but let
me describe the initial effort in this direction and see how far
we can go. What we will do here is to construct a model that
possesses correct symmetries (e.g., chiral symmetry etc.) and
density dependence that describes the ground state of nuclear
matter and then fluctuate around that ground state to compute
interesting response functions.
\subsubsection{Walecka mean field theory with intrinsic density
dependence}
 \itt
Let us consider that $\Phi_L$ now contains nucleons, pions and
vector mesons. In fact the same can be formulated with the vector
mesons and other heavy mesons integrated out but it is convenient
to put the vectors and a scalar meson denoted $\phi$ explicitly.
In free space this scalar may be the scalar meson with a broad
width which is being hotly debated  nowadays. For the symmetric
nuclear matter for which we will apply mean field approximation,
we can drop the pions and the $\rho$ mesons. The Lagrangian we
will consider is
 \be
\calL=\bar{N}(i\gamma_\mu (\del^\mu+ig^\star \omega^\mu)-M^\star
+h^\star\phi)N-\frac 14 F^2_{\mu\nu}+\frac 12 {m^\star_\omega}^2
\omega^2 +\frac 12 (\del_\mu\phi)^2 -\frac 12 {m^\star_\phi}^2
\phi^2.\label{walecka*}
 \ee
Apart from the stars appearing in the parameters, this is of the
same form as Walecka's linear mean field model [SEROT]. I should
emphasize that this Lagrangian can be made consistent with chiral
symmetry by restoring pion fields and identifying $\phi$ as a
chiral scalar, not as the fourth component of the chiral four
vector as in linear sigma model. The $\omega$ meson is a chiral
scalar.

Suppose now that we do the mean field approximation in computing
nuclear matter properties. If one simply puts the density
dependence as $m^\star=m(n)$, $n$ being the matter density, and do
the mean field, one does not get the right answer. In fact, one
loses the conservation of energy-momentum tensor and
thermodynamics does not come out right as a consequence. This has
been a long standing difficulty in naively putting density
dependence into the parameters of a mean field theory Lagrangian.

There may be more than one resolutions to this problem. One such
solution was found by Song, Min and Rho [SMR,SONG]. The idea is to
make the parameters of the Lagrangian a chiral invariant
functional of the density field $operator$ $\check{n}$ defined by
 \be
\check{n} u^\mu=\bar{N}\gamma^\mu N \label{densityop}
 \ee
with unit fluid 4-velocity
 \be
u^\mu=\frac{1}{\sqrt{1-\vec{v}^2}}(1,\vec{v})=\frac{1}{\sqrt{n^2-
\vec{j}^2}} (n,\vec{j}).
 \ee
Here $\vec{j}=\la \bar{N}\vec{\gamma} N\ra$ is the baryon current
density and $n\equiv \la N^\dagger N\ra$ is the baryon number
density. In mean field, the field dependence brings additional
terms in the baryon equation of motion whereas it does not affect
the meson equations of motion. These additional terms contribute
to the energy momentum tensor a term of the form
 \be
\delta T^{\mu\nu}=-2 \check{\Sigma}\bar{N}\gamma_\mu n^\mu N
g^{\mu\nu}
 \ee
where
 \be
\check{\Sigma}=\frac{\del \calL}{\del\check{n}}
 \ee
which comes into the energy density and the pressure of the
system. In many-body theory language, this additional term at mean
field order corresponds to ``rearrangement" terms. This simple
manipulation restores all the conservation laws lost when a
c-number density is used. The resulting energy density etc. is of
the same form as that of linear Walecka model except that the mass
and coupling constants are a function of density. Let me just use
the scaling consistent with the BR scaling
 \be
\frac{m_\omega^\star}{m_\omega}\approx
\frac{m_\phi^\star}{m_\phi}\approx\frac{M^\star}{M}=\Phi
(n)\label{scaling1}
 \ee
and
 \be
\frac{g^\star}{g}\approx \Phi (n)\label{scaling2}
 \ee
with a universal scaling factor
 \be
\Phi (n)=\frac{1}{1+0.28 n/n_0}.
 \ee
The numerical value 0.28 in the denominator of $\Phi$ will be seen
coming from nuclear gyromagnetic ratio in lead nuclei. Using the
standard value for the masses for the $\omega$ and the nucleon and
$m_\phi\approx 700$ MeV, I get the binding energy $BE=16.0$ MeV,
the equilibrium density $k_{eq}=257.3$ MeV, $m^\star/M=0.62$ and
the compression modulus $K=296$ MeV. These results agree well with
experiments, much better than the simple Walecka linear model, in
particular for the compression modulus. In fact they are
comparable to the ``best-fit" mean field model [FURNST] which
gives $BE=16.0\pm 0.1$ MeV, $k_{eq}=256\pm 2$ MeV,
$m^\star/M=0.61\pm 0.03$ and $K=250\pm 50$ MeV.

The use of the bilinear nucleon field operator (\ref{densityop})
makes evident and direct the dependence on density of the
parameters of the Lagrangian in the mean field approximation. One
could perhaps use other field variables such as the scalar field
$\phi$ which is chiral scalar. However the scalar field would
affect only the equation of motion of the scalar field, not that
of the fermion field. It would make the density dependence
complicated since the ground-state expectation value (GEV) of the
scalar field is only indirectly related to the number density.
Also doing the mean field with the scalar would involve different
approximations. This is an interesting possibility to explore,
however; it has not yet been studied.

One more point to note. Use of the scaling relations
(\ref{scaling1}) and (\ref{scaling2}) is an additional ingredient
to the notion that there should be an intrinsic dependence on
density in the parameters of the Lagrangian. It is not dictated by
the RGEs and the QCD-HLS matching.
\subsubsection{Response functions of a quasiparticle}
 \itt
Given the ground state as described above, we now would like to
calculate the response to an external field of a quasiparticle
sitting on top of the Fermi sea. Here I will discuss response to
the electromagnetic (EM) field. The response to the weak field in
the axial channel is not yet well understood.

Consider a (non-relativistic) quasiparticle sitting on top of the
Fermi sea with the momentum $\vec{p}$ which is probed by a slowly
varying EM field. The convection current is given by
 \be
\vec{J}=g_l \frac{\vec{p}}{M}\label{convection}
 \ee
with $|\vec{p}|\approx k_F$, where $M$ is the free-space nucleon
mass and $g_l$ is the orbital gyromagnetic ratio given by
 \be
g_l=\frac{1+\tau_3}{2}+\delta g_l.
 \ee
It is important to note that it is the free-space mass $M$, not an
effective mass $m^\star$, that appears in (\ref{convection}). This
is so because of the charge conservation or more generally gauge
invariance. In condensed matter physics, such a requirement is
known for the cyclotron frequency of an electron in magnetic field
as ``Kohn theorem."

To calculate $g_l$, one starts with a chiral Lagrangian with the
parameters of the Lagrangian density-dependent as determined by
the mean field property as described in the last subsection and
then calculates fluctuations on top of the ground state. This has
been discussed extensively in the literature [FRS:98, RHO:MIG,
BR:PR01], so I shall not go into details. The calculation is
rather straightforward and the result is
 \be
\delta g_l= \frac 49\left[\Phi^{-1} -1-\frac 12
\tilde{F}^\pi_1\right]\tau_3\label{deltagl}
 \ee
where $\tilde{F}^\pi_1$ is the $l=1$ component of the Landau
parameter $F$ -- the spin- and isospin-independent component of
the quasiparticle interaction $\calF$ -- coming from one-pion
exchange which is of course given unambiguously by the chiral
Lagrangian for a given $k_F$. The expression is valid for density
up to $\sim n_0$ but cannot be pushed to the regime where $\Phi\ll
1$. (When $\Phi\ll 1$, then the nucleon cannot be treated
non-relativistically and hence (\ref{deltagl}) will break down.)

For nuclear matter density,
 \be
\tilde{F}^\pi_1 (n_0)=-0.153
 \ee
and from an experiment for a proton in the lead region
 \be
\delta g_l^p=0.23\pm 0.03.
 \ee
This then gives
 \be
\Phi (n_0)\approx 0.78.
 \ee
In terms of the parameterization $\Phi (n)=(1+y n/n_0)^{-1}$, this
corresponds to $y=0.28$ used before for the ground state of
nuclear matter.

Other applications of the relation are discussed in the references
given above. What we have gotten here is a beginning of the
relation between the hadronic parameters effective in medium of an
effective Lagrangian matched to QCD and many-body interactions
characterized by a number of fixed points around the Fermi
surface. Much work needs to be done to unravel the intricate
connections that we have a glimpse of.
\section{Comments on the Literature}
\itt I summarize the main references I used for my lectures. They
will be incomplete, since I focus mainly only on issues I have
been working on recently.  I will be leaving out many of the
important papers in the field for which my apologies.
 \vskip 0.3cm
$\bullet$ {\bf Lecture I}:\vskip 0.3cm

The idea of Cheshire Cat was first introduced formally by
Nadkarni, Nielsen and Zahed~\cite{NNZ:CC} and phenomenologically
by Brown, Jackson, Rho and Vento~\cite{BJRV}. The subsequent
developments are summarized in \cite{RHO:PR,NRZ} on which my
discussions are based. Some of the similar topics are reviewed by
Hosaka and Toki~\cite{HOSAKA}. The resolution of ``proton spin
problem" in the context of Cheshire Cat is described in
\cite{LMPRV,HJL}.
 \vskip 0.3cm

$\bullet$ {\bf Lecture II: Dilute matter}:\vskip 0.3cm

An early attempt to incorporate Weinberg's counting
rule~\cite{WEIN:CHI} into nuclear chiral effective field theory
was made in 1982~\cite{RHO:ERICE} . It was incomplete, however,
and it was only after Weinberg's 1990 paper on the counting rule
in pion-nuclear interactions that what was missing in 1982 was
restored~\cite{RHO:PRL}. Applying Weinberg's counting rule to
nuclear force was made by Ord\'o$\tilde{\rm n}$ez and van
Kolck~\cite{OVK}. Since then, many papers have been published in
which counting schemes different from that of Weinberg have been
investigated. These developments are extensively summarized in
\cite{BEANE}. The importance of conformal invariant fixed point in
nuclear physics was pointed out by Mehen, Stewart and
Wise~\cite{MEHEN}. That the Weinberg counting rule would not be
invalidated if one were to do a regularization appropriate to EFT
in the Wilsonian sense (i.e., the role of power divergences in
EFT) was argued in \cite{PKMR:CO}. There have been published a
large number of papers that improve on the construction of
two-nucleon as well as multi-nucleon forces to high orders in
chiral perturbation theory. The most recent efforts are summarized
in \cite{EPEL1,EPEL2}.

In \cite{PKMR:TAI}, the thesis was put forward that the power of
EFT in nuclear physics lies in making predictions for processes
that are important for other areas of physics (such as
astrophysics) but can be provided neither by the standard nuclear
physics approach alone nor by the straightforward low-order power
counting approach. This point was illustrated in the precise
calculation of the solar $pp$ process~\cite{PMSV:pp} and $hep$
process~\cite{PMSV:hep}. In doing these calculations, it was
argued that incorporating accurate nuclear wave functions obtained
in the SNPA into the framework of EFT as an integral part of EFT
is both justified and more effective~\footnote{I like to call this
``more effective effective field theory (MEEFT)" since it exploits
both the wealth of information acquired by the practitioners of
SNPA and the consistency with EFT. In fact, I challenged the
aficionados of the ``strict-counting rule" to come up with a
prediction of the $hep$ process without unknown parameters. If
they can get, say, within a year, a prediction which does better
than the result of [PMSV:hep], I will offer a bottle of Premier
Grand Cru Ch\^ateau Mouton-Rothschild.} than the approach based on
SNPA alone or on strict adherence to counting rules.\vskip 0.3cm

$\bullet$ {\bf Lecture II: Superdense matter} \vskip 0.3cm

The literature on this topic is huge. Since color
superconductivity per se is not the main focus of this part of
lecture, I use only the review by Rajagopal and Wilczek~\cite{RW}.
One can find an extensive list of references in this review. The
EQCD (effective QCD) Lagrangian obtained at high density by
decimating toward the Fermi surface was first made (as far as I
know) by Hong~\cite{HONG}. The first formulation as an EFT of the
collective modes that arise in the color-flavor locking was made
simultaneously by Hong, Rho and Zahed~\cite{HRZ} (hep-ph/9903503)
and by Casalbuoni and Gatto~\cite{CG} (hep-ph/9908227). It was
argued in \cite{HRZ} that solitons in the effective chiral
Lagrangian in the CFL phase, called ``qualitons," could be
identified as baryons in the system. It was realized then that the
EFT Lagrangian with the gluons included is of the HLS form and
that bound-state excitations of diquarks could arise in the
super-pion and super-vector-meson channels [RWZ]. These could be
coupled to the Golstone bosons and Higgsed vector mesons in a way
analogous to the ``sobar" description [KRBR]. The ``quark-hadron
continuity" conjecture was put forward by Sch\"afer and
Wilczek~\cite{SW:CONT}.\vskip 0.3cm

$\bullet$ {\bf Lecture III}: \vskip 0.3cm

The idea that the color-flavor locking ``observed" at superhigh
density can also take place at zero density was presented by
Wetterich in a series of papers~\cite{WETT} soon after the CFL was
proposed at high density. That Wetterich's CFL scenario -- which
is a ``top-down" approach -- coincides with Harada-Yamawaki's HLS
scenario -- which is a ``bottom-up" approach -- was noticed in
\cite{BR:BERK,BR:PR01}. The discovery of vector manifestation in
chiral symmetry by Harada and
Yamawaki~\cite{HY:VM,HY:FATE,HY:MATCH,HY:PR} led to a series of
new developments discussed in the lecture, namely the vanishing of
the vector meson mass at the chiral
transition~\cite{HY:CONF,HS:T,HKR} and the derivation of one of BR
scalings. The interplay of Wetterich's approach and
Harada-Yamawaki's approach suggests an equivalence of explicit and
hidden gauge symmetries, somewhat analogous to Weinberg's
observation~\cite{WEIN:GAUGE} on non-uniqueness in going from one
to the other between non-gauge-symmetric effective field theory
and gauge symmetric theory. \vskip 0.3cm

$\bullet$ {\bf Lecture IV}:\vskip 0.3cm

The notion that Landau Fermi liquid theory is a bona-fide EFT has
been advocated and developed by, among others,
Shankar~\cite{SHANK}, Polchinski~\cite{POL} and Fr\"olich et
al~\cite{CFROS}. That Walecka theory of nuclear matter is
equivalent to Landau-Migdal Fermi liquid theory was shown sometime
ago by Matsui~\cite{MATSUI}. By relating BR scaling to
Fermi-liquid fixed point parameters, the authors in
\cite{FR:96,FRS:98,SMR,SONG} constructed an effective field theory
for nuclear matter in terms of BR scaling parameters. This mapping
of chiral Lagrangian field theory to Fermi-liquid fixed point
theory has met with success in explaining the anomalous
gyromagnetic ratio in heavy nuclei~\cite{FRS:98} but has not been
satisfactorily tested in weak axial transitions other than
axial-charge transitions. How to incorporate axial responses in
the framework of Landau theory and effective chiral Lagrangian is
not yet worked out. The present status in this direction is
summarized in \cite{RHO:MIG}.

Arriving at nuclear matter starting with chiral Lagrangians
defined in matter-free space was initiated seriously by
Lynn~\cite{LYNN}. Recent works focusing on chiral perturbation
theory in medium posit the presence of a Fermi sea and develop
perturbations around the Fermi surface~\cite{KFW}.


\subsection*{Acknowledgments}
\itt I have benefitted from recent discussions and/or
collaborations with Gerry Brown, Masayasu Harada, Youngman Kim,
Vicente Vento and Koichi Yamawaki. Giving these lectures at the
spectacularly beautiful mountain region Hualien, Taiwan was a real
pleasure. I would like to thank the organizers, in particular Shin
Nan Yang and Wen-Chen Chang, of the Spring School for the
opportunity. Some of these lectures were prepared when I was
visiting KIAS in the Autumn 2001 and I would like to acknowledge
the Institute's hospitality.

\newpage

\end{document}